# Mechanically Verified Calculational Abstract Interpretation

David Darais
University of Maryland
davdar@cs.umd.edu

David Van Horn
University of Maryland
dvanhorn@cs.umd.edu

## Abstract

Calculational abstract interpretation, long advocated by Cousot, is a technique for deriving correct-by-construction abstract interpreters from the formal semantics of programming languages.

This paper addresses the problem of deriving correct-by-*verified*-construction abstract interpreters with the use of a proof assistant. We identify several technical challenges to overcome with the aim of supporting verified calculational abstract interpretation that is faithful to existing pencil-and-paper proofs, supports calculation with Galois connections generally, and enables the extraction of verified static analyzers from these proofs.

To meet these challenges, we develop a theory of Galois connections in monadic style that include a *specification effect*. Effectful calculations may reason classically, while pure calculations have extractable computational content. Moving between the worlds of specification and implementation is enabled by our metatheory.

To validate our approach, we give the first mechanically verified proof of correctness for Cousot's "Calculational design of a generic abstract interpreter." Our proof "by calculus" closely follows the original paper-and-pencil proof and supports the extraction of a verified static analyzer.

## 1. Introduction

Abstract interpretation [9, 10] is a foundational and unifying theory of semantics and abstraction developed by P. Cousot and R. Cousot, which has had notable impact on the theory and practice of program analysis and verification. Traditionally, static analyses and verification frameworks such as type systems, program logics, or constraint-based analyses start by first postulating a specification of an abstract semantics. Only afterward is this abstraction proved correct with respect to the language's semantics. This proof establishes *post facto* that the analysis or logic *is* an abstract interpretation of the underlying language semantics.

P. Cousot has also advocated an alternative approach to the design of abstract interpreters called *calculational* abstract interpretation [7, 8], which involves systematically applying abstraction functions to a programming language semantics in order to *derive* an abstraction. Abstract interpretations derived in the calculational style are correct by construction (assuming no missteps are made in the calculation) and need not be proved sound after the fact.

This paper addresses the problem of mechanically verifying the derivations of calculational abstract interpretation using a proof assistant. We identify several technical challenges to modelling the theory of abstract interpretation in a constructive, dependent type theory and then develop solutions to these challenges. Paramount in overcoming these challenges is effectively representing Galois connections and maintaining a modality between specifications and implementations to enable program extraction. To do this, we propose a novel form of Galois connections endowed with monadic structure which we dub *Kleisli Galois connections*. This monadic structure maintains a distinction between calculation at the specification level, which may be non-constructive, and at the implementation level, which must be constructive. Remarkably, calculations are able to move back and forth between these modalities and verified programs may be extracted from the end result of calculation.

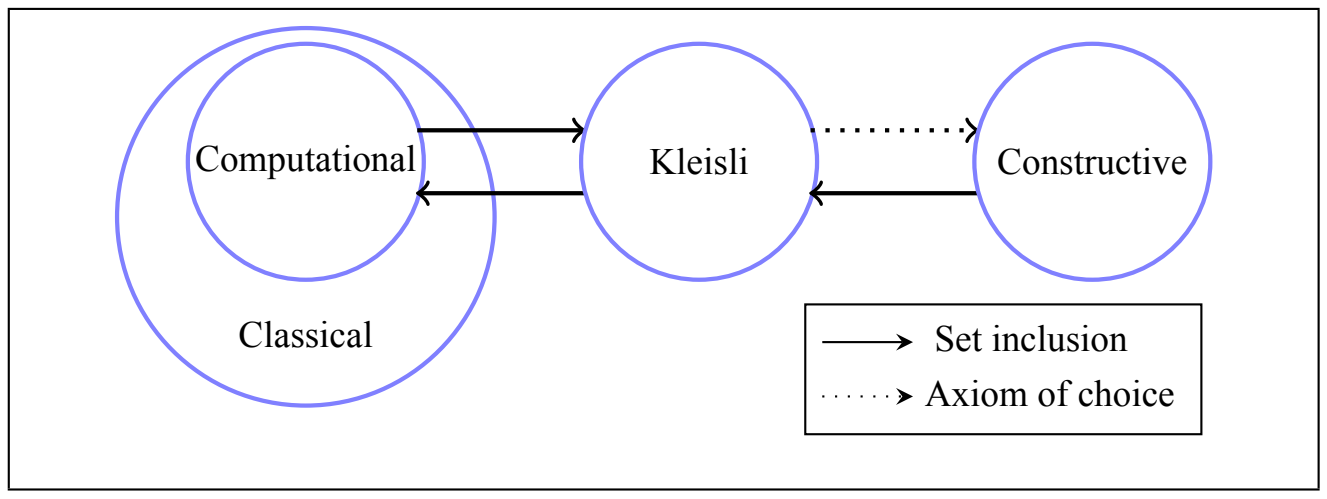

**Figure 1:** Relations between classical Galois connections and their Kleisli and constructive counterparts.

To establish the adequacy of our theory, we prove it is sound and complete with respect to a subset of traditional Galois connections, and isomorphic to a space of fully constructive Galois connections, diagrammed in figure 1. To establish the utility of our theory, we construct a framework for abstract interpretation with Kleisli Galois connections in the dependently typed programming language and proof-assistant, Agda [20]. To validate our method, we re-derive Cousot's generic compositional static analyzer for an imperative language by abstract interpretation of the language's formal semantics. Consequently we obtain a verified proof of the calculation and extract a verified implementation of Cousot's static analyzer.

***Contributions*** This paper contributes:

1. a framework for mechanically verified abstract interpretation that supports calculation and program extraction,
2. a theory of specification effects in Galois connections, and
3. a verified proof of Cousot's generic abstract interpreter derived by calculus.

To supplement these contributions, we provide two artifacts. The first is the source code of this document, which is a literate Agda program and verified at typesetting-time. For presentation purposes, it assumes a few lemmas and is less general than it could be. The second artifact is a stand-alone Agda program that develops all of the results in this paper in full detail, including the mathematically stated theorems and lemmas. Claims are marked with a "✓" whenever they have been proved in Agda. (All claims are checked.) The full development is found at:

`https://github.com/plum-umd/mvcai`

Although largely self-contained, this paper assumes a basic familiarity with abstract interpretation and dependently typed programming. There are excellent tutorials on both ([8, 11] and [5, 21], respectively).

## 2. Calculational Abstract Interpretation

To demonstrate our approach to mechanizing Galois connections we present the calculation of a generic abstract interpreter, as originally presented by Cousot [7]. The setup is a simple arithmetic expression language which includes a random number expression, and is otherwise standard. The syntax and semantics is given in figure 2.

A collecting semantics is defined as a monotonic (written $\nearrow$) predicate transformer using $_ \vdash _ \mapsto _$:

$$eval \in exp \to \wp(env) \nearrow \wp(val)$$
$$eval[e](R) := \{v \mid \exists \rho \in R : \rho \vdash e \mapsto v\}$$

In the setting of abstract interpretation, an analysis for a program $e$ is performed by: (1) defining another semantics $eval^\sharp$, where $eval^\sharp$ is shown to soundly approximate the semantics of *eval*, and (2) executing the $eval^\sharp[e]$ semantics and observing the output. There are many different methods for arriving at $eval^\sharp$, however the calculational approach prescribes a methodology for defining $eval^\sharp$ through calculus, the results of which are correct by construction.

To arrive at $eval^\sharp$ through calculus we fist establish an abstraction for the domain $\wp(env) \nearrow \wp(val)$, which we call $env^\sharp \nearrow val^\sharp$. After abstracting the domain, we induce a *best specification* for any abstract semantics $eval^\sharp \in env^\sharp \nearrow val^\sharp$. Then we perform calculation on this specification to arrive at a *definition* for $eval^\sharp$. Key in this methodology is the requirement that $eval^\sharp$ be an *algorithm*, otherwise we would just define $eval^\sharp$ to be the induced best specification and be done.

We induce the best specification for *eval* by: (1) constructing an abstraction for values $val^\sharp$ and proving it is a valid abstraction of $\wp(val)$, (2) constructing an abstraction for environments $env^\sharp$ and proving it is a valid abstraction of $\wp(env)$, (3) lifting these abstractions pointwise to $env^\sharp \nearrow val^\sharp$ and proving it is a valid abstraction of $\wp(env) \nearrow \wp(val)$, and (4) inducing $\alpha^{e\to v}(eval)$ as the best abstraction of *eval* using the results from (3).

***Abstracting values*** We pick a simple sign abstraction for $val^\sharp$, however our final calculated abstract interpreter will be fully generic to $val^\sharp$, as is done in Cousot's original derivation [7].

$$v^\sharp \in val^\sharp := \{-, 0, +, \top, \bot\}$$

The set $val^\sharp$ has the partial ordering $\bot \sqsubseteq - \parallel 0 \parallel + \sqsubseteq \top$ where $_ \parallel _$ is notation for incomparable.

Justifying that $val^\sharp$ is a valid abstraction takes the form of a Galois connection:

$$\wp(val) \underset{\alpha^v}{\overset{\gamma^v}{\leftrightarrows}} val^\sharp.$$

Galois connections are mappings between concrete objects and abstract objects which satisfy soundness and completeness properties. For $val^\sharp$, the Galois connection with $\wp(val)$ is defined:

$$\begin{aligned} &\alpha^v \in \wp(val) \nearrow val^\sharp & &\gamma^v \in val^\sharp \nearrow \wp(val) \\ &\alpha^v(V) := -\ \text{if}\ \exists v \in V : v < 0 & &\gamma^v(-) := \{v \mid v < 0\} \\ &\qquad \sqcup\ 0\ \ \text{if}\ 0 \in V & &\gamma^v(0) := \{0\} \\ &\qquad \sqcup\ +\ \text{if}\ \exists v \in V : v > 0 & &\gamma^v(+) := \{v \mid v > 0\} \\ & & &\gamma^v(\top) := \mathbb{Z} \\ & & &\gamma^v(\bot) := \varnothing \end{aligned}$$

$\alpha^v$ is called the *abstraction* function, which maps concrete sets of numbers in $\wp(val)$ to a finite, symbolic representations in $val^\sharp$. $\gamma^v$ is called the *concretization* function, mapping abstract symbols in $val^\sharp$ concrete sets in $\wp(val)$.

This Galois connection is *extensive*: properties of values in $val^\sharp$ imply properties of related concrete values in $\wp(val)$. It is *reductive*: $\alpha^v$ is the best possible abstraction given $\gamma^v$.

| | | | | | |
|---|---|---|---|---|---|
| $n$ | $\in$ | *lit* | $:=$ | $\mathbb{Z}$ | integer literals |
| $x$ | $\in$ | *var* | | | variables |
| $u$ | $\in$ | *unary* | $::=$ | $+ \mid -$ | unary operators |
| $b$ | $\in$ | *binary* | $::=$ | $+ \mid - \mid \times \mid \%$ | binary operator |
| $e$ | $\in$ | *exp* | $::=$ | $n$ | integer literal |
| | | | $\mid$ | $x$ | variable |
| | | | $\mid$ | *rand* | random integer |
| | | | $\mid$ | $u\ e$ | unary operator |
| | | | $\mid$ | $e\ b\ e$ | binary operator |

| | |
|---|---|
| $v \in val := \mathbb{Z}$ | values |
| $[\![_]\!]^u \in unary \to (val \to val)$ | unary op denot. |
| $[\![_]\!]^b \in binary \to (val \times val \rightharpoonup val)$ | binary op denot. |
| $\rho \in env := var \rightharpoonup val$ | environments |
| $_ \vdash _ \mapsto _ \in \wp(env \times exp \times val)$ | eval. relation |

$$\frac{}{\rho \vdash n \mapsto n} \qquad \frac{}{\rho \vdash x \mapsto \rho(x)} \qquad \frac{n \in \mathbb{Z}}{\rho \vdash rand \mapsto n}$$

$$\frac{\rho \vdash e \mapsto v}{\rho \vdash u\ e \mapsto [\![u]\!]^u(v)} \qquad \frac{\rho \vdash e_1 \mapsto v_1 \qquad \rho \vdash e_2 \mapsto v_2}{\rho \vdash e_1\ b\ e_2 \mapsto [\![b]\!]^b(v_1, v_2)}$$

**Figure 2:** Syntax and semantics

**Lemma 1** (*extensive*$^v$).✓ *$\gamma^v \circ \alpha^v$ is extensive, that is:*

$$\forall (V \in \wp(val)).V \subseteq \gamma^v(\alpha^v(V)).$$

**Lemma 2** (*reductive*$^v$).✓ *$\alpha^v \circ \gamma^v$ is reductive, that is:*

$$\forall (v^\sharp : val^\sharp).\alpha^v(\gamma^v(v^\sharp)) \sqsubseteq v^\sharp.$$

***Abstracting environments*** We abstract $\wp(env)$ with $env^\sharp$:

$$\rho^\sharp \in env^\sharp := var \rightharpoonup val^\sharp$$

Justifying that $env^\sharp$ is a valid abstraction is done through a Galois connection $\wp(env) \underset{\alpha^e}{\overset{\gamma^e}{\leftrightarrows}} env^\sharp$:

$$\alpha^e : \wp(env) \nearrow env^\sharp$$
$$\alpha^e(R) := \lambda(x).\alpha^v(\{\rho(x) \mid \rho \in R\})$$
$$\gamma^e : env^\sharp \nearrow \wp(env)$$
$$\gamma^e(\rho^\sharp) := \{\rho \mid \forall (x).\rho(x) \in \gamma^v(\rho^\sharp(x))\}$$

**Lemma 3** (*extensive*$^e$).✓ *$\alpha^e \circ \gamma^e$ is extensive, that is:*

$$\forall (R \in \wp(env)).R \subseteq \gamma^e(\alpha^e(R)).$$

**Lemma 4** (*reductive*$^e$).✓ *$\gamma^e \circ \alpha^e$ is reductive, that is:*

$$\forall (\rho^\sharp \in env^\sharp).\alpha^e(\gamma^e(\rho^\sharp)) \sqsubseteq \rho^\sharp.$$

***Abstracting the function space*** To abstract $\wp(env) \nearrow \wp(val)$, we abstract its components pointwise with $env^\sharp \nearrow val^\sharp$, and justify the abstraction with another Galois connection.

$$\alpha^{e\to v} : (\wp(env) \nearrow \wp(val)) \nearrow (env^\sharp \nearrow val^\sharp)$$
$$\alpha^{e\to v}(f) := \alpha^v \circ f \circ \gamma^e$$
$$\gamma^{e\to v} : (env^\sharp \nearrow val^\sharp) \nearrow (\wp(env) \nearrow \wp(val))$$
$$\gamma^{e\to v}(f^\sharp) := \gamma^v \circ f^\sharp \circ \alpha^e$$

**Lemma 5** (*extensive*$^{e\to v}$).✓ *$\gamma^{e\to v} \circ \alpha^{e\to v}$ is extensive, that is:*

$$\forall (f \in \wp(env) \nearrow \wp(val)).f \sqsubseteq \gamma^{e\to v}(\alpha^{e\to v}(f)).$$

**Numeric literals**

| | |
|---|---|
| $\alpha^v(\{v \mid \exists \rho \in \gamma^r(\rho^\sharp) : \rho \vdash n \mapsto v\})$ | |
| $= \alpha^v(\{n\})$ | ⟅ definition of $\rho \vdash n \mapsto v$ ⟆ |
| $\triangleq eval^\sharp[n](\rho^\sharp)$ | ⟅ by defining $eval^\sharp[n](\rho^\sharp) \coloneqq \alpha^v(\{n\})$ ⟆ |

**Variable Reference**

| | |
|---|---|
| $\alpha^v(\{v \mid \exists \rho \in \gamma^r(\rho^\sharp) : \rho \vdash x \mapsto v\})$ | |
| $= \alpha^v(\{\rho(x) \mid \rho \in \gamma^r(\rho^\sharp)\})$ | ⟅ definition of $\rho \vdash x \mapsto v$ ⟆ |
| $= \alpha^{e \to v}(\lambda(R).\{\rho(x) \mid \rho \in R\})(\rho^\sharp)$ | ⟅ definition of $\alpha^{e \to v}$ ⟆ |
| $\sqsubseteq \rho^\sharp(x)$ | ⟅ Fact: $\alpha^{e \to v}(\lambda(\rho).\{\rho(x) \mid \rho \in R\}) \sqsubseteq (\lambda(\rho^\sharp).\rho^\sharp(x))$ ⟆ |
| $\triangleq eval^\sharp[x](\rho^\sharp)$ | ⟅ by defining $eval^\sharp[x](\rho^\sharp) \coloneqq \rho^\sharp(x)$ ⟆ |

**Unary operators**

| | |
|---|---|
| $\alpha^v(v \mid \exists \rho \in \gamma^r(\rho^{)} : \rho \vdash u\ e \mapsto v)$ | |
| $= \alpha^v(\{[\![u]\!]^u(v) \mid \exists \rho \in \gamma^r(\rho^\sharp) : \rho \vdash e \mapsto v\})$ | ⟅ definition of $\rho \vdash u\ e \mapsto v$ ⟆ |
| $\sqsubseteq \alpha^v(\{[\![u]\!]^u(v) \mid v \in \gamma^v(\alpha^v(\{v' \mid \exists \rho \in \gamma^r(\rho^\sharp) : \rho \vdash e \mapsto v\}))\})$ | ⟅ $\alpha^v$ monotone and $\gamma^v \circ \alpha^v$ extensive ⟆ |
| $= \alpha^v(\{[\![u]\!]^u(v) \mid v \in \gamma^v(\alpha^{e \to v}(eval[e])(\rho^\sharp))\})$ | ⟅ definition of $\alpha^{e \to v}$ and $eval[e]$ ⟆ |
| $\sqsubseteq \alpha^v(\{[\![u]\!]^u(v) \mid v \in \gamma^v(eval^\sharp[e](\rho^\sharp))\})$ | ⟅ $\alpha^v$ monotonic and inductive hypothesis for $e$ ⟆ |
| $= \alpha^{v \to v}(\lambda(V).\{[\![u]\!]^u(v) \mid v \in V\})(eval^\sharp[e](\rho^\sharp))$ | ⟅ definition of $\alpha^{v \to v}$ ⟆ |
| $\sqsubseteq [\![u]\!]^{u\sharp}(eval^\sharp[e](\rho^\sharp))$ | ⟅ by assuming $\alpha^{v \to v}(\lambda(V) \to \{[\![u]\!]^u(v) \mid v \in V\}) \sqsubseteq [\![u]\!]^{u\sharp}$ ⟆ |
| $\triangleq eval^\sharp[u\ e](\rho^\sharp)$ | ⟅ by defining $eval^\sharp[u\ e](\rho^\sharp) \coloneqq [\![u]\!]^{u\sharp}(eval^\sharp[e](\rho^\sharp))$ ⟆ |

**Figure 3:** Cousot's *classical* calculation of the Generic Abstract Interpreter

**Lemma 6** ($reductive^{e \to v}$).✓ *$\alpha^{e \to v} \circ \gamma^{e \to v}$ is reductive, that is:*

$$\forall (f^\sharp \in env^\sharp \nrightarrow val^\sharp).\alpha^{e \to v}(\gamma^{e \to v}(f^\sharp)) \sqsubseteq f^\sharp.$$

***Inducing a best specification*** A best specification for any abstraction of *eval* is induced from the Galois connection

$$\wp(env) \nrightarrow \wp(val) \underset{\alpha^{e \to v}}{\overset{\gamma^{e \to v}}{\leftrightarrows}} env^\sharp \nrightarrow val^\sharp$$

as $\alpha^{e \to v}(eval)$, or $\alpha^v \circ eval \circ \gamma^e$. An abstract semantics $eval^\sharp$ can then be shown to satisfy this specification through an ordered relationship $\alpha^{e \to v}(eval) \sqsubseteq eval^\sharp$.

The process of calculation is to construct $eval^\sharp$ through a chain of ordered reasoning: $\alpha^{e \to v}(eval[e]) = \alpha^v \circ eval[e] \circ \gamma^e \sqsubseteq ... \sqsubseteq eval^\sharp[e]$ such that $eval^\sharp$ is an algorithm, at which point we have defined $eval^\sharp[e]$ through calculation.

### 2.1 Calculating the Abstract Interpreter

The calculation of $eval^\sharp$ begins by expanding definitions:

| | |
|---|---|
| $\alpha^{e \to v}(eval[e])(\rho^\sharp)$ | |
| $= \alpha^v(eval[e](\gamma^r(\rho^\sharp)))$ | ⟅ definition of $\alpha^{e \to v}$ ⟆ |
| $= \alpha^v(\{v \mid \exists \rho \in \gamma^r(\rho^\sharp) : \rho \vdash e \mapsto v\})$ | ⟅ definition of $eval[e]$ ⟆ |

In case $\gamma^r(\rho^\sharp) = \emptyset$, then have $\alpha^{e \to v}(eval[e])(\rho^\sharp) = \alpha^v(\emptyset) = \bot$. Otherwise, we proceed by induction on $e$, assuming $\gamma^r(\rho^\sharp)$ is nonempty.

In figure 3, we show the calculations for literals, variables, and unary operator expressions. This calculation is generic, meaning it is parameterized by implementations for abstracting random numbers, and unary and binary operators. The parameters for the unary operator case are an abstract unary denotation $[\![_]\!]^{u\sharp} \in val^\sharp \nrightarrow val^\sharp$ and a proof that it abstracts concrete unary denotation:

$$\alpha^{v \to v}(\lambda(V).\{[\![u]\!]^u(v) \mid v \in V\})(v^\sharp) \sqsubseteq [\![u]\!]^{u\sharp}(v^\sharp)$$

The calculation for the remaining forms can be found in Cousot's notes [8, lec. 16]. This calculation serves to contrast the constructive calculation we develop in section 4.5, which is more amenable to verification and extraction in Agda.

## 3. Mechanization: The Easy Parts

We aim to mechanize calculations of the style presented in figure 3. Some parts are easy; we start with those.

Figure 4 gives the syntax and semantics in Agda. Variables are modelled as an index into an environment of statically known size; otherwise, the syntax of *exp* translates directly. The meaning of unary operators is given by a function, ⟦_⟧ᵘ, while binary operators are defined relationally, _ ∈ ⟦_⟧ᵇ_, to account for the partiality of [/] and [%], which take elements of $\mathbb{Z}^+$: integers paired with a proof of being non-zero. Environments are modelled intentionally as a list of values, rather than extensionally with Agda's function space. Environments are statically well-formed to contain a value mapping for every variable, resulting in a total lookup function _[_]. Partiality is eliminated from the definition of environment lookup by static well-formedness. The relational semantics is defined using a dependent inductive type, _ ⊢ _ ↦ _.

To encode *eval*, we create powersets using *characteristic functions*, assuming set-theoretic primitives (defined later), where the judgement $x$ ∈ mk[℘](φ) holds *iff* φ(x) is inhabited.

```
℘      : Set → Set
mk[℘]  : ∀ {A} → (A → Set) → ℘ A
_∈_    : ∀ {A} → A → ℘ A → Set
```

The eval function is then defined using an existential type inside of a characteristic function:

```
eval[_] : ∀ {Γ} → exp Γ → ℘ (env Γ) → ℘ val
eval[ e ] R = mk[℘] (λ v → ∃ ρ st (ρ ∈ R) × (ρ ⊢ e ↦ v))
```

## 4. Constructive Abstract Interpretation

The Galois connections presented in the section 2 are not immediately amenable to encoding in Agda, or constructive logic in general. The heart of the problem is the definition of $\alpha^v$:

$$\begin{array}{rcl} \alpha^v(V) & \coloneqq & -\text{ if } \exists v \in V : v < 0 \\ & \sqcup & 0\ \text{ if } 0 \in V \\ & \sqcup & +\text{ if } \exists v \in V : v > 0 \end{array}$$

```
val = ℤ
size = ℕ
data var : size → Set where
  Zero : ∀ {Γ} → var (Suc Γ)
  Suc  : ∀ {Γ} → var Γ → var (Suc Γ)

data unary  : Set where [+] [−] : unary
data binary : Set where [+] [−] [×] [/] [%] : binary

data exp (Γ : size) : Set where
  Num       : ℤ → exp Γ
  Var       : var Γ → exp Γ
  Rand      : exp Γ
  Unary[_]  : unary → exp Γ → exp Γ
  Binary[_] : binary → exp Γ → exp Γ → exp Γ

⟦_⟧^u : unary → val → val
⟦ [+] ⟧^u = suc
⟦ [−] ⟧^u = pred

data _∈⟦_⟧^b_ : val → binary → val × val → Set where
  [+] : ∀ {v1 v2} →                   (v1 + v2)            ∈⟦ [+] ⟧^b (v1 , v2)
  [−] : ∀ {v1 v2} →                   (v1 − v2)            ∈⟦ [−] ⟧^b (v1 , v2)
  [×] : ∀ {v1 v2} →                   (v1 × v2)            ∈⟦ [×] ⟧^b (v1 , v2)
  [/] : ∀ {v1 v2}
        → (p : v2 ≢ Zero) → (v1 / mk[ℤ^+] v2 p) ∈⟦ [/] ⟧^b (v1 , v2)
  [%] : ∀ {v1 v2}
        → (p : v2 ≢ Zero) → (v1 % mk[ℤ^+] v2 p) ∈⟦ [%] ⟧^b (v1 , v2)

data env : size → Set where
  [] : env Zero
  _::_ : ∀ {Γ} → val → env Γ → env (Suc Γ)

_[_] : ∀ {Γ} → env Γ → var Γ → val
(v :: ρ) [ Zero  ] = v
(v :: ρ) [ Suc x ] = ρ [ x ]

data _⊢_↦_ {Γ} : env Γ → exp Γ → val → Set where
  Num    : ∀ {ρ n}                 → ρ ⊢ Num n             ↦ n
  Var    : ∀ {ρ x}                 → ρ ⊢ Var x             ↦ ρ [ x ]
  Rand   : ∀ {ρ n}                 → ρ ⊢ Rand              ↦ n
  Unary  : ∀ {ρ o e v1 v2}
           → v2 ≡ ⟦ o ⟧^u v1        → ρ ⊢ e                 ↦ v1
                                    → ρ ⊢ Unary[ o ] e      ↦ v2
  Binary : ∀ {ρ o e1 e2 v1 v2 v3}
           → v3 ∈⟦ o ⟧^b (v1 , v2)  → ρ ⊢ e1                ↦ v1
                                    → ρ ⊢ e2                ↦ v2
                                    → ρ ⊢ Binary[ o ] e1 e2 ↦ v3
```

**Figure 4:** Syntax and semantics in Agda

A literal translation of $\alpha^v$ to constructive logic would require *deciding* predicates such as $\exists v \in V : v < 0$ in order to return a value of type $val^\sharp$, however such predicates are in general undecidable.

There are a number of known options for encoding $\alpha^v$, each of which has shortcomings for our goal of extracting computation from the result of a verified calculation.

***Non-solution 1: Admit Excluded Middle*** One option to defining $\alpha^v$ is to to postulate the law of excluded middle:

```
excluded−middle : ∀ (P : Set) → P ⊎ (¬ P)
```

This axiom imbues the logic with classical reasoning, is logically consistent, and would allow us to perform case analysis on the existential predicate $\exists v \in V : v < 0$ to complete a definition for $\alpha^v$. This approach has the drawback that definitions no longer carry computational content, and cannot be extracted or computed with in general.

***Non-solution 2: Work in Powerset*** Another option is to always work inside the powerset type $\wp$, meaning $\alpha^v$ would have type $\wp(val) \nearrow \wp(val^\sharp)$. This approach also allows for a successful definition of $\alpha^v$, but again suffers from not being a computation. Functions at type $\wp(val) \nearrow \wp(val^\sharp)$ cannot be executed to produce values at type $val^\sharp$, which is the end goal.

***Non-solution 3: Only use Concretization*** The state of the art in mechanizing abstract interpreters is to use "$\gamma$-only" encodings of soundness and completeness properties [14]. However, this approach has a number of drawbacks: it does not support calculation, it gives the engineer no guidance as to whether or not their $\gamma$ is sensible (sound and complete w.r.t. $\alpha$), and it is less compositional than the Galois connection framework.

### 4.1 Our Solution: A Specification Effect

The problem of encoding Galois connections in constructive logic exists with an apparent dichotomy: if the construction is too classical then one cannot extract computation from the result, and if it is too constructive it prevents the definition of classical structures like Galois connections. We find a solution to this problem through a new Galois connection framework which marks the transition from constructive to classical with a *monadic effect*. Classical and constructive reasoning can then be combined within the same framework, and classical constructions can be promoted to constructive ones after they are shown to be effect-free.

We find a solution to the problem of encoding calculational abstract interpretation in constructive logic by reformulating the definition of a Galois connection into the powerset Kleisli category. This approach:

1. is rooted in the first principles of Galois connections,
2. allows for the definition of Galois connections which would otherwise require classical reasoning (like excluded middle),
3. supports abstract interpretation by calculus, and
4. allows for the extraction of algorithms from calculations.

The transition to the powerset Kleisli category results in abstraction and concretization mappings which have a *specification effect*, meaning they return a classical powerset value, which we model non-constructively. The laws that accompany the Galois connection will then introduce and eliminate this effect. Combined with monad laws, which also introduce and eliminate monadic effects, we are able to mix constructive and classical reasoning and extract an algorithm from the result of calculation, after all introduced effects have been eliminated.

### 4.2 Kleisli Galois Connections

Kleisli Galois connections are formed by re-targeting the classical Galois connection framework from the category of posets to the powerset Kleisli category. The morphisms in this category are monotonic monadic functions $A \nearrow \wp(B)$ rather than their classical counterparts $A \nearrow B$. Powersets $\wp(A)$ are required to be monotonic themselves, meaning they are *downward closed*, i.e. $X \in \wp(A)$ is monotonic if and only if $\forall (x, y). x \in X \to y \sqsubseteq x \to y \in X$.

The reflexive morphism in the powerset Kleisli category is *return*, rather than *id*, where *return* is defined as the downward

**Numeric literals**

$$
\begin{array}{ll}
\alpha^{mv} * (\mathit{eval}^m[n] * (\gamma^{mr}(\rho^\sharp))) & \\
= \alpha^{mv} * (\{v \mid \exists \rho \in \gamma^{mr}(\rho^\sharp) : \rho \vdash n \mapsto v\}) & \wr \text{ definition of } \mathit{eval}^m[n]* \int \\
\sqsubseteq \alpha^{mv} * (\mathit{return}(n)) & \wr \text{ definition of } \rho \vdash n \mapsto v \int \\
= \alpha^{mv}(n) & \wr \text{ monad right unit } \int \\
= \mathit{return}(\eta^v(n)) & \wr \text{ induced } \eta^v \text{ from } \alpha^{mv} \int \\
\triangleq \mathit{return}(\mathit{eval}^{m\sharp}[n](\rho^\sharp)) & \wr \text{ by defining } \mathit{eval}^{m\sharp}[n](\rho^\sharp) := \eta^v(n) \int
\end{array}
$$

**Variable references**

$$
\begin{array}{ll}
\alpha^{mv} * (\mathit{eval}^m[x] * (\gamma^{mr}(\rho^\sharp))) & \\
= \alpha^{mv} * (\{v \mid \exists \rho \in \gamma^{mr}(\rho^\sharp) : \rho \vdash x \mapsto v\}) & \wr \text{ definition of } \mathit{eval}^m[x]* \int \\
= \alpha^{mv} * (\{\rho(x) \mid \rho \in \gamma^{mr}(\rho^\sharp)\}) & \wr \text{ definition of } \rho \vdash x \mapsto v \int \\
= \alpha^{mv} * ((\lambda(\rho).\mathit{return}(\rho(x))) * (\gamma^{mr}(\rho^\sharp))) & \wr \text{ monad unit and associativity } \int \\
= \alpha^{e \to^m v}(\lambda(\rho).\mathit{return}(\rho(x)))(\rho^\sharp) & \wr \text{ definition of } \alpha^{e \to^m v} \int \\
\sqsubseteq \mathit{return}(\rho^\sharp(x)) & \wr \text{ Fact: } \alpha^{e \to^m v}(\lambda(\rho).\mathit{return}(\rho(x))) \sqsubseteq (\lambda(\rho^\sharp).\mathit{return}(\rho^\sharp(x))) \int \\
\triangleq \mathit{return}(\mathit{eval}^{m\sharp}[x](\rho^\sharp)) & \wr \text{ by defining } \mathit{eval}^{m\sharp}[x](\rho^\sharp) := \rho^\sharp[x] \int
\end{array}
$$

**Unary operators**

$$
\begin{array}{ll}
\alpha^{mv} * (\mathit{eval}^m[u\ e] * (\gamma^{mr}(\rho^\sharp))) & \\
= \alpha^{mv} * (\{[\![u]\!]^u(v) \mid \exists \rho \in \gamma^{mr}(\rho^\sharp) : \rho \vdash e \mapsto v\}) & \wr \text{ definition of } \mathit{eval}^m[u\ e]* \int \\
= \alpha^{mv} * ((\lambda(v).\mathit{return}([\![u]\!]^u(v))) * (\{v \mid \exists \rho \in \gamma^{mr}(\rho^\sharp) : \rho \vdash e \mapsto v\})) & \wr \text{ monad unit and associativity } \int \\
\sqsubseteq \alpha^{mv} * ((\lambda(v).\mathit{return}([\![u]\!]^u(v)))* & \\
\quad (\{v \mid v \in \gamma^{mv} * (\alpha^{mv} * (\{v' \mid \exists \rho \in \gamma^{mr}(\rho^\sharp) : \rho \vdash e \mapsto v'\}))\})) & \wr\ \gamma^{mv} \diamond \alpha^{mv} \text{ extensive } \int \\
= \alpha^{mv} * ((\lambda(v).\mathit{return}([\![u]\!]^u(v))) * (\{v \mid v \in \gamma^{mv} * (\alpha^{e \to^m v}(\mathit{eval}^m[e])(\rho^\sharp))\})) & \wr \text{ definition of } \alpha^{e \to^m v} \text{ and } \mathit{eval}^m[e] \int \\
\sqsubseteq \alpha^{mv} * ((\lambda(v).\mathit{return}([\![u]\!]^u(v))) * (\gamma^{mv} * (\mathit{return}(\mathit{eval}^{m\sharp}[e](\rho^\sharp))))) & \wr \text{ monotonicity of } \alpha^v, \mathit{return} \text{ and } *, \text{ and IH for } e \int \\
= \alpha^{v \to^m v}(\lambda(v).\mathit{return}([\![u]\!]^u(v))) * (\mathit{return}(\mathit{eval}^{m\sharp}[e](\rho^\sharp))) & \wr \text{ definition of } \alpha^{v \to^m v} \text{ and monad associativity } \int \\
\sqsubseteq (\lambda(v^\sharp).\mathit{return}([\![u]\!]^{u\sharp}(v^\sharp))) * (\mathit{return}(\mathit{eval}^{m\sharp}[e](\rho^\sharp))) & \wr \text{ by assuming } \alpha^{v \to^m v}(\lambda(v).\mathit{return}([\![u]\!]^u(v)))(v^\sharp) \sqsubseteq \mathit{return}([\![u]\!]^{u\sharp}(v^\sharp)) \int \\
= \mathit{return}([\![u]\!]^{u\sharp}(\mathit{eval}^{m\sharp}[e](\rho^\sharp))) & \wr \text{ monad right unit } \int \\
\triangleq \mathit{return}(\mathit{eval}^{m\sharp}[u\ e])(\rho^\sharp) & \wr \text{ by defining } \mathit{eval}^{m\sharp}[u\ e](\rho^\sharp) := [\![u]\!]^u(\mathit{eval}^{m\sharp}(\rho^\sharp)) \int
\end{array}
$$

**Figure 5:** Our *constructive* calculation of the Generic Abstract Interpreter

closure of the singleton set:

$$
\begin{aligned}
&\mathit{return} \in \forall(A).A \nearrow \wp(A) \\
&\mathit{return}(x) = \{y \mid y \sqsubseteq x\}
\end{aligned}
$$

The monadic bind operator, which we call *extension* and notate _* in the tradition of Moggi [18], is defined using a dependent sum, or existential type:

$$
\begin{aligned}
&_* \in \forall(A, B).(A \nearrow \wp(B)) \nearrow (\wp(A) \nearrow \wp(B)) \\
&f * (X) = \{y \mid \exists x \in X : y \in f(x)\}
\end{aligned}
$$

To establish that $\wp$ forms a monad with *return* and _* we prove left-unit, right-unit and associativity laws.

**Lemma 7** ($\wp$-monad)**.**✓ *$\wp$ forms a monad with return and _*, meaning the following properties hold:*

$$
\begin{aligned}
\mathit{left\text{-}unit} &: \forall(X).\mathit{return} * (X) = X \\
\mathit{right\text{-}unit} &: \forall(f, x).f * (\mathit{return}(x)) = f(x) \\
\mathit{associativity} &: \forall(f, g, X).g * (f * (X)) = (\lambda(x).g * (f(x))) * (X)
\end{aligned}
$$

Composition in the powerset Kleisli category is notated _ $\diamond$ _ and defined with _*:

$$
\begin{aligned}
&_ \diamond _ \in \forall(A, B, C).(B \nearrow \wp(C)) \to (A \nearrow \wp(B)) \to A \nearrow \wp(C) \\
&(g \diamond f)(x) = g * (f(x))
\end{aligned}
$$

A Kleisli Galois connection $A \underset{\alpha^m}{\overset{\gamma^m}{\leftrightarrows}} B$, which we always notate with superscripts $\alpha^m$ and $\gamma^m$, is analogous to that of classical Galois connection but with monadic morphisms, unit and composition:

$$
\begin{aligned}
&\alpha^m \in A \nearrow \wp(B) \\
&\gamma^m \in B \nearrow \wp(A) \\
&\mathit{extensive}^m : \forall(x).\mathit{return}(x) \sqsubseteq \gamma^m * (\alpha^m(x)) \\
&\mathit{reductive}^m : \forall(x^\sharp).\alpha^m * (\gamma^m(x^\sharp)) \sqsubseteq \mathit{return}(x^\sharp)
\end{aligned}
$$

The presence of *return* as the identity is significant: *return* marks the transition from pure values to those which have a "specification effect". *extensive*$^m$ states that $\gamma^m \diamond \alpha^m$ is a pure computation *at best*, and *reductive*$^m$ states that $\alpha^m \diamond \gamma$ is a pure computation *at worst*. The consequence of this will be important during calculation: appealing to *extensive*$^m$ and *reductive*$^m$ will introduce and eliminate the specification effect, respectively.

### 4.3 Lifting Kleisli Galois Connections

The end goal of our calculation is stated as a partial order relationship using a classical Galois connection: $\alpha^{e \to v}(\mathit{eval}) \sqsubseteq \mathit{eval}^\sharp$. If we wish to work with Kleisli Galois connections, we must build bridges between Kleisli results and classical ones. This bridge is stated as an isomorphism between Kleisli Galois connections and a subset of classical Galois connections that hold computational content, as shown in section 1 figure 1. In addition to the Galois connections themselves, we map proofs of relatedness between Kleisli and classical Galois connections, so long as the classical result is of the form $\alpha^{A \to B}(f*) \sqsubseteq f^\sharp*$ where $f$ and $f^\sharp$ are monadic functions.

In order to leverage Kleisli Galois connections for our calculation we must recognize *eval* as the extension of a monotonic monadic function $eval^m$. Recall the definition of *eval*:

$$eval[e] \in \wp(env) \nearrow \wp(val)$$
$$eval[e](R) := \{v \mid \exists \rho \in R : \rho \vdash e \mapsto v\}$$

This is the extension of the monadic powerset function: $eval^m$:

$$eval^m[e] \in env \nearrow \wp(val)$$
$$eval^m[e](\rho) := \{v \mid \rho \vdash e \mapsto v\}$$

where, by definition of $_ {*}$:

$$eval^m[e] * (R) = \{v \mid \exists \rho \in R : \rho \vdash e \mapsto v\} = eval[e](R)$$

This observation enables us to construct a Kleisli Galois connection:

$$env \to \wp(val) \underset{\alpha^{e \to^m v}}{\overset{\gamma^{e \to^m v}}{\leftrightarrows}} env^\sharp \to \wp(val^\sharp)$$

and calculation

$$\alpha^{e \to^m v}(eval[e]) \sqsubseteq eval^\sharp[e],$$

and lift the results to classical ones automatically via the *soundness* of the mapping from Kleisli to classical. Furthermore, we know that any classical Galois connection and classical calculation of $eval^\sharp$ can be encoded as Kleisli via the *completeness* of the Kleisli to classical mapping. We give precise definitions for *soundness* and *completeness* in section 7.

### 4.4 Constructive Galois Connections

When performing calculation to discover $eval^{m\sharp}[e]$ from the induced specification $\alpha^{e \to^m v}(eval^m[e])$, we will require that the result be an algorithm, which we can now state precisely as having no monadic effect. The goal will then be to calculate the pure function $eval^{m\sharp}[e] \in env^\sharp \nearrow val^\sharp$ such that

$$\alpha^{e \to^m v}(eval^m[e])(\rho^\sharp) \sqsubseteq return(eval^{m\sharp}[e](\rho^\sharp))$$

However, at present, such a calculation will be problematic. Take for instance, the definition we would like to end up with for numeric literal expressions:

$$eval^{m\sharp}[n](\rho^\sharp) := \alpha^{mv}(n)$$

This defines the abstract interpretation of a numeric literal as the immediate lifting of that literal to an abstract value. However this definition is not valid, since $\alpha^{mv} \in val \nearrow \wp(val^\sharp)$ introduces a specification effect. The problem becomes magnified when we wish to parameterize over $\alpha^{mv}$, as Cousot does in his original derivation.

One idea is to restrict all $\alpha^m$ mappings to be pure, and only parameterize over abstractions for *val* which have pure mappings. We take morally this approach, although later we show that it is not a restriction at all, and arises naturally through an isomorphism between Kleisli Galois connections and those which have pure abstraction functions, which we call *constructive Galois connections*. This isomorphism is visualized on the right-hand-side of figure 1, and proofs are given in section 7.

A constructive Galois connection is a variant of Kleisli Galois connection where the abstraction function $\alpha^m$ is required to have no specification effect, which we call $\eta$ following the convention of [19, p. 237] where it is called an "extraction function":

$$\eta : A \nearrow B$$
$$\gamma^c : B \nearrow \wp(A)$$
$$extensive^c : return(x) \sqsubseteq \gamma * return(\eta(x))$$
$$reductive^c : (\lambda(x).return(\eta(x))) * \gamma(x^\sharp) \sqsubseteq return(x^\sharp)$$

We can now define the abstract interpretation for numeric literals as:

$$eval^{m\sharp}[n](\rho^\sharp) := \eta^v(n)$$

which is a pure computation that can be extracted and executed.

### 4.5 Calculating the Interpreter, Constructively

We now recast Cousot's calculational derivation of a generic abstract interpreter in the setting of Kleisli Galois connections. In the next section we show how the constructive version is translatable to Agda. As before, we only show numeric literals, variable reference and unary operators; see our full Agda development for constructive calculations of the remaining cases.

Recall the constructive calculation goal, which is to discover a *pure* function $eval^{m\sharp}$ such that

$$\alpha^{e \to^m v}(eval^m)(\rho^\sharp) \sqsubseteq return(eval^{m\sharp}(\rho^\sharp))$$

This goal makes it clear that we are starting with a specification $eval^m : env \nearrow \wp(val)$, and working towards a pure computation $eval^{m\sharp} : env^\sharp \nearrow val^\sharp$. The process of calculation will eliminate the "specification effect" from the induced specification $\alpha^{e \to^m v}(eval^m)$ using monad laws and the reductive and expansive properties of Kleisli Galois connections.

The setting assumes Kleisli Galois connections for the abstractions of values $val \underset{\alpha^{mv}}{\overset{\gamma^{mv}}{\leftrightarrows}} val^\sharp$, environments $env \underset{\alpha^{me}}{\overset{\gamma^{me}}{\leftrightarrows}} env^\sharp$ and their induced classical Galois connection for the monadic function space $val \nearrow \wp(env) \underset{\alpha^{e \to^m v}}{\overset{\gamma^{e \to^m v}}{\leftrightarrows}} val^\sharp \nearrow \wp(env^\sharp)$. When needed we replace $\alpha^m(x)$ with an equivalent pure extraction function $return(\eta(x))$ using the isomorphism between Kleisli and constructive Galois connections.

We begin calculating from the specification $\alpha^{e \to^m v}(eval^m)$ by unfolding definitions:

$$\begin{aligned} &\alpha^{e \to^m v}(eval^m[e])(\rho^\sharp) \\ &= (\alpha^{mv} \diamond eval^m[e] \diamond \gamma^{mr})(\rho^\sharp) && \wr \text{ definition of } \alpha^{e \to^m v} \text{ ∫} \\ &= \alpha^{mv} * (eval^m[e] * (\gamma^{mr}(\rho^\sharp))) && \wr \text{ monad associtivity ∫} \end{aligned}$$

and proceed by induction on *e*. The calculations for numeric literals, variables and unary operators are show in figure 4. The parameters for the unary operator case in the constructive setting are an abstract unary denotation $[\![_]\!]^{u\sharp} \in val^\sharp \nearrow val^\sharp$ and a proof that it abstracts concrete unary denotation:

$$\alpha^{v \to^m v}(\lambda(v).return([\![u]\!]^u(v)))(v^\sharp) \sqsubseteq return([\![u]\!]^{u\sharp}(v^\sharp))$$

The biggest difference between our constructive derivation and Cousot's classical derivation is the presence of monadic unit *return* and extension operator $_ {*}$. In the process of calculation, monadic unit and associativity laws are used in combination with extensive and reductive properties to calculate toward a pure value, at which point the result is both a pure computation and an abstraction of *eval*[*e*] simultaneously by construction.

## 5. Galois Connection Metatheory in Agda

We now encode our constructive calculation of the Generic Abstract Interpreter in Agda, both to verify the results mechanically and to extract an executable version of the resulting abstract interpreter.

We mechanize the calculation of the interpreter first by developing a theory of posets, its monotonic function space, and a non-constructive powerset type, which we prove is a monad. Then we develop theories of classical, Kleisli and constructive Galois connections, as well as their soundness and completeness relationships. Finally, we embed the constructive calculation in Agda, arriving at at an executable algorithm, and lift its correctness property to the classical correctness criteria initially specified by Cousot.

### 5.1 Posets in Agda

We begin by defining PreOrd, a relation which is reflexive and transitive. PreOrd is a *type class*, meaning top-level instance definitions will be automatically selected by Agda during type inference.

```
record PreOrd (A : Set) : Set where
  field
    _≤_ : A → A → Set
    xRx⌜≤⌝ : ∀ {x} → x ≤ x
    _⊖⌜≤⌝_ : ∀ {x y z} → y ≤ z → x ≤ y → x ≤ z
```

We then define posets in Agda:

```
data POSet : Set where
  ⇧ : (A : Set) → {{PO : PreOrd A}} → POSet
```

The double curly brackets around PO indicate that the argument should be inferred through *type class resolution*, which we rely on heavily in our development.

To construct a POSet (rather than the builtin Set), we create another datatype 《_》, which selects the domain of a POSet.

```
dom : POSet → Set
dom (⇧ A {{PO}}) = A
```

```
data 《_》 (A : POSet) : Set where
  ↑⟨_⟩ : dom A → 《 A 》
```

The reason for introducing a new datatype 《_》 is purely technical; it allows us to block reduction of elements of ⇧ $A$ until we witness its lifting from a value $x$ : dom $A$ into ↑⟨ $x$ ⟩ : POSet $A$.

Next, we induce a partial order on POSet from the antisymmetric closure of the supplied pre order lifted to elements of 《_》:

```
[_]_≤⌜dom⌝_ : ∀ (A : POSet) → dom A → dom A → Set
[ ⇧ A {{PO}} ] x ≤⌜dom⌝ y = x ≤ y
```

```
data _⊑_ {A : POSet} : 《 A 》 → 《 A 》 → Set where
  ↑⟨_⟩ : ∀ {x y : dom A} → [ A ] x ≤⌜dom⌝ y → ↑⟨ x ⟩ ⊑ ↑⟨ y ⟩
```

This definition of _ ⊑ _ is designed to also block reduction until the liftings of $x$ and $y$ are likewise witnessed through pattern matching.

We induce equivalence as the antisymmetric closure of _ ⊑ _.

```
data _≈_ {A : POSet} (x y : 《 A 》) : Set where
  _,_ : x ⊑ y → y ⊑ x → x ≈ y
```

We prove reflexivity, transitivity and antisymmetry for _ ⊑ _, as well as reflexivity, transitivity and symmetry for _ ≈ _:

```
xRx⌜⊑⌝ : ∀ {A : POSet} {x : 《 A 》} → x ⊑ x
_⊖⌜⊑⌝_ : ∀ {A : POSet} {x y z : 《 A 》} → y ⊑ z → x ⊑ y → x ⊑ z
⋈⌜≈⌝ : ∀ {A : POSet} {x y : 《 A 》} → x ⊑ y → y ⊑ x → x ≈ y
xRx⌜≈⌝ : ∀ {A : POSet} {x y : 《 A 》} → x ≈ y → x ⊑ y
_⊖⌜≈⌝_ : ∀ {A : POSet} {x y z : 《 A 》} → y ≈ z → x ≈ y → x ≈ z
◦⌜≈⌝ : ∀ {A : POSet} {x y : 《 A 》} → x ≈ y → y ≈ x
```

Now we can define the two most important posets: the function space and powerset.

### 5.2 Monotonic Functions in Agda

To construct a poset for monotonic functions we carry proofs of monotonicity around with each function.

```
data mon (A B : POSet) : Set where
  mk[mon] : (f : 《 A 》 → 《 B 》) →
    {f−proper : ∀ {x y} → x ⊑ y → f x ⊑ f y} → mon A B
```

The PreOrd for mon is the pointwise ordering of ⊑:

```
data _≤⌜mon⌝_ {A B : POSet} : mon A B → mon A B → Set where
  ↑⟨_⟩ : ∀ {f : 《 A 》 → 《 B 》} {f−proper : ∀ {x y} → x ⊑ y → f x ⊑ f y}
    {g : 《 A 》 → 《 B 》} {g−proper : ∀ {x y} → x ⊑ y → g x ⊑ g y}
  → (∀ {x} → f x ⊑ g x)
  → mk[mon] f {f−proper} ≤⌜mon⌝ mk[mon] g {g−proper}
```

We lift mon to a POSet with _ ⇒ _:

```
_⇒_ : POSet → POSet → POSet
A ⇒ B = ⇧ (mon A B)
```

Application in _ ⇒ _ is _ · _:

```
_·_ : ∀ {A B : POSet} → 《 A ⇒ B 》 → 《 A 》 → 《 B 》
↑⟨ mk[mon] f {f−proper} ⟩ · x = f x
```

We define a helper function for creating values in _ ⇒ _ which require no monotonicity proof (which we use for demonstration purposes only):

```
mk[⇒] : ∀ {A B : POSet} → (《 A 》 → 《 B 》) → 《 A ⇒ B 》
mk[⇒] f = ↑⟨ mk[mon] f {f−proper} ⟩ where
  postulate f−proper : ∀ {x y} → x ⊑ y → f x ⊑ f y
```

For example, composition is defined as _ ⊙ _:

```
_⊙_ : ∀ {A B C : POSet} → 《 B ⇒ C 》 → 《 A ⇒ B 》 → 《 A ⇒ C 》
g ⊙ f = mk[⇒] (λ x → g · (f · x))
```

### 5.3 Monotonic Powerset in Agda

We define powersets as monotonic characteristic functions into Agda's Set type.

```
data pow (A : POSet) : Set where
  mk[pow] : (φ : 《 A 》 → Set) →
    {φ−proper : ∀ {x y} → y ⊑ x → φ x → φ y} → pow A
```

Whereas mk[⇒] $f$ {*f-proper*} constructs a monotonic function from $f$, mk[℘] $\varphi$ {*φ-proper*} constructs a set from a monotonic characteristic function $\varphi$. Antitonicity of the argument to $\varphi$ in the statement of *φ-proper* is required to ensure sets are downward closed.

The preorder for pow is implication:

```
data _≤⌜pow⌝_ {A : POSet} : pow A → pow A → Set where
  ↑⟨_⟩ : ∀ {φ₁ : 《 A 》 → Set} {φ₁−proper : ∀ {x y} → y ⊑ x → φ₁ x → φ₁ y}
    {φ₂ : 《 A 》 → Set} {φ₂−proper : ∀ {x y} → y ⊑ x → φ₂ x → φ₂ y}
  → (∀ {x} → φ₁ x → φ₂ x)
  → mk[pow] φ₁ {φ₁−proper} ≤⌜pow⌝ mk[pow] φ₂ {φ₂−proper}
```

We lift pow into the POSet type as ℘.

```
℘ : POSet → POSet
℘ A = ⇧ (pow A)
```

The set-containment judgement is _ ∈ _.

```
_∈_ : ∀ {A : POSet} → 《 A 》 → 《 ℘ A 》 → Set
x ∈ ↑⟨ mk[pow] φ {φ−proper} ⟩ = φ x
```

And like functions we provide a cheat for creating monotonic sets without the burden of monotonicity proof.

```
mk[℘] : ∀ {A : POSet} → (《 A 》 → Set) → 《 ℘ A 》
mk[℘] φ = ↑⟨ mk[pow] φ {φ−proper} ⟩ where
  postulate φ−proper : ∀ {x y} → y ⊑ x → φ x → φ y
```

```
record _⇄_ (A B : POSet) : Set where -- Classical
  field
    α[_] : ⟪ A ⇒ B ⟫
    γ[_] : ⟪ B ⇒ A ⟫
    extensive[_] : ∀ {x : ⟪ A ⟫}
                 → x ⊑ γ[_] · (α[_] · x)
    reductive[_] : ∀ {x♯ : ⟪ B ⟫}
                 → α[_] · (γ[_] · x♯) ⊑ x♯

record _⇄ᵐ_ (A B : POSet) : Set where -- Kleisli
  field
    αᵐ[_] : ⟪ A ⇒ ℘ B ⟫
    γᵐ[_] : ⟪ B ⇒ ℘ A ⟫
    extensiveᵐ[_] : ∀ {x : ⟪ A ⟫}
                  → return · x ⊑ γᵐ[_] * · (αᵐ[_] · x)
    reductiveᵐ[_] : ∀ {x♯ : ⟪ B ⟫}
                  → αᵐ[_] * · (γᵐ[_] · x♯) ⊑ return · x♯

record _⇄ᶜ_ (A B : POSet) : Set where -- Constructive
  field
    η[_] : ⟪ A ⇒ B ⟫
    γᶜ[_] : ⟪ B ⇒ ℘ A ⟫
    extensiveᶜ[_] : ∀ {x : ⟪ A ⟫}
                  → return · x ⊑ γᶜ[_] * · (pure · η[_] · x)
    reductiveᶜ[_] : ∀ {x♯ : ⟪ B ⟫}
                  → (pure · η[_]) * · (γᶜ[_] · x♯) ⊑ return · x♯
```

**Figure 6:** Classical, Kleisli, and constructive Galois connections.

### 5.4 Powerset Monad in Agda

The powerset monad has unit return, where return $x$ is the set of all elements smaller than $x$, as defined by a characteristic function:

```
return : ∀ {A : POSet} → ⟪ A ⇒ ℘ A ⟫
return = mk[⇒] (λ x → mk[℘] (λ y → y ⊑ x))
```

We lift the return operation to functions, which we call pure.

```
pure : ∀ {A B : POSet} → ⟪ (A ⇒ B) ⇒ A ⇒ ℘ B ⟫
pure = mk[⇒] (λ f → mk[⇒] (λ x → return · (f · x)))
```

Monadic extension is _*.

```
_* : ∀ {A B : POSet} → ⟪ A ⇒ ℘ B ⟫ → ⟪ ℘ A ⇒ ℘ B ⟫
f * = mk[⇒] (λ X → mk[℘] (λ y → ∃ x st (x ∈ X) × y ∈ f · x))
```

We use _* to define Kleisli composition, _⋄_:

```
_⋄_ : ∀ {A B C : POSet} → ⟪ B ⇒ ℘ C ⟫ → ⟪ A ⇒ ℘ B ⟫ → ⟪ A ⇒ ℘ C ⟫
g ⋄ f = mk[⇒] (λ x → g * · (f · x))
```

Finally, we prove (and omit) monads laws analogous to those in lemma 7.

## 6. Calculational Abstract Interpretation in Agda

We show Agda types for classical, Kleisli and constructive Galois connections in figure 6. Using these definitions we calculate an abstract interpreter in Agda following the constructive approach described in section 4 in the following steps:

1. Define a constructive Galois connection between *env* and $env^\sharp$.
2. Lift (1) and a parameterized Galois connection for *val* pointwise to the monotonic function space.
3. Induce the specification for an abstraction of a monadic semantic function $eval^m[e]$ as $\alpha^{e \to^m v}(eval^m[e])$.
4. Calculate over $\alpha^{e \to^m v}(eval^m[e])$ until we arrive at a pure function $pure(eval^{m\sharp}[e])$.
5. Lift the relationship $\alpha^{e \to^m v}(eval^m[e]) \sqsubseteq pure(eval^{m\sharp}[e])$ to the classical abstraction of function extensions $\alpha^{(e \to^m v)*}(eval^m[e]*) \sqsubseteq pure(eval^{m\sharp}[e])*$ through a mechanized proof of soundness of Kleisli Galois connections w.r.t. classical.

### 6.1 Abstracting Environments in Agda

We define a *constructive* Galois connection between env and env♯ rather than Kleisli because it results in nicer definitions. First we parameterize by an abstraction for values, which we do with an Agda module:

```
module §–env♯ (val♯ : POSet) (⇄val⇄ : ⇧ val ⇄ᶜ val♯) where
```

Abstract environments take the form of another list-like inductive datatype:

```
data env♯ : size → Set where
  [] : env♯ Zero
  _::_ : ∀ {Γ} → ⟪ val♯ ⟫ → env♯ Γ → env♯ (Suc Γ)

_[_]♯ : ∀ {Γ} → env♯ Γ → var Γ → ⟪ val♯ ⟫
(v♯ :: ρ) [ Zero ]♯ = v♯
(v♯ :: ρ) [ Suc x ]♯ = ρ [ x ]♯
```

The ordering for env♯ is the pointwise ordering:

```
data _≤ᵉ_ : ∀ {Γ} → env♯ Γ → env♯ Γ → Set where
  [] : [] ≤ᵉ []
  _::_ : ∀ {Γ} {ρ₁ ρ₂ : env♯ Γ} {v₁ v₂}
    → v₁ ⊑ v₂ → ρ₁ ≤ᵉ ρ₂ → (v₁ :: ρ₁) ≤ᵉ (v₂ :: ρ₂)
```

The environment abstraction function $\eta^e$ is the pointwise application of η[ ⇄ val ⇄ ]:

```
ηᵉ : ∀ {Γ} → env Γ → env♯ Γ
ηᵉ [] = []
ηᵉ (n :: ρ) = η[ ⇄val⇄ ] · ↑⟨ n ⟩ :: ηᵉ ρ
```

The concretization function $\gamma^{ce}$ is the pointwise concretization of γᶜ[ ⇄ val ⇄ ]:

```
data _∈γᵉ_ : ∀ {Γ} → env Γ → env♯ Γ → Set where
  [] : [] ∈γᵉ []
  _::_ : ∀ {Γ} {ρ : env Γ} {ρ♯ : env♯ Γ} {n n♯}
    → ↑⟨ n ⟩ ∈ γᶜ[ ⇄val⇄ ] · n♯ → ρ ∈γᵉ ρ♯ → (n :: ρ) ∈γᵉ (n♯ :: ρ♯)
```

The $\eta^e$ and _ ∈ γᵉ_ functions are sound and complete by pointwise applications of soundness and completness from ⇄ val ⇄:

```
soundᶜᵉ : ∀ {Γ} (ρ : env Γ) → ρ ∈γᵉ ηᵉ ρ
soundᶜᵉ [] = []
soundᶜᵉ (x :: ρ) = soundᶜ[ ⇄val⇄ ] :: soundᶜᵉ ρ

completeᶜᵉ : ∀ {Γ} {ρ : env Γ} {ρ♯} → ρ ∈γᵉ ρ♯ → ηᵉ ρ ≤ ρ♯
completeᶜᵉ [] = []
completeᶜᵉ (n∈γ[n♯] :: ρ∈γ[ρ♯]) =
  completeᶜ[ ⇄val⇄ ] n∈γ[n♯] :: completeᶜᵉ ρ∈γ[ρ♯]
```

From these definitions, we construct ⇄ env ⇄ : ∀ {Γ} → ⇧ (env Γ) η ⇄ γ ⇧ (env$^\sharp$ Γ) using a helper function mk[⇄$^c$ ↑] for lifting primitive definitions (non-POSet) to Galois connections.

```
⇄env⇄ : ∀ {Γ} → ⇧ (env Γ) ⇄^c ⇧ (env^♯ Γ)
⇄env⇄ = mk[⇄^c↑] η^e (λ ρ^♯ ρ → ρ ∈γ^e ρ^♯) sound^ce complete^ce
```

### 6.2 Inducing a Best Specification in Agda

The monadic semantics is encoded with the evaluation relation:

```
eval^m[_] : ∀ {Γ} → exp Γ → ⟪ ⇧ (env Γ) ⇒ ℘ (⇧ val) ⟫
eval^m[ e ] = mk[↑⇒] (λ ρ → mk[℘↑] (λ v → ρ ⊢ e ↦ v))
```

To induce a best abstraction we first encode the pointwise lifting of two Kleisli Galois connections ⇄ $A$ ⇄ and ⇄ $B$ ⇄ to classical pointwise Galois connections over the monadic function space as (⇄ $A$ ⇄ ⇒ ⌜⇄$^m$⌝ ⇄ $B$ ⇄).

```
_⇒⌜⇄^m⌝_ : ∀ {A₁ A₂ B₁ B₂ : POSet}
           → A₁ ⇄^m A₂ → B₁ ⇄^m B₂ → (A₁ ⇒ ℘ B₁) ⇄ (A₂ ⇒ ℘ B₂)
_⇒⌜⇄^m⌝_ {A₁} {A₂} {B₁} {B₂} ⇄A⇄ ⇄B⇄ = record
  { α[_] = mk[⇒] (λ f → α^m[ ⇄B⇄ ] ⋄ f ⋄ γ^m[ ⇄A⇄ ])
  ; γ[_] = mk[⇒] (λ f^♯ → γ^m[ ⇄B⇄ ] ⋄ f^♯ ⋄ α^m[ ⇄A⇄ ])
  ; extensive[_] = … ; reductive[_] = … }
```

We can now state the specification for eval$^m$[ $e$ ] as a pure function which approximating α[ ⇄ env ⇄⇒ ⌜⇄$^c$⌝ ⇄ val ⇄ ] · eval$^m$[ $e$ ].

### 6.3 Calculating the Interpreter, in Agda

Before calculating we must lift the various semantic functions to the monotonic function space, like ⟦_⟧$^u$, _[_] and _[_]$^\sharp$:

```
↑⟦_⟧^u : unary → ⟪ ⇧ val ⇒ ⇧ val ⟫
lookup[_] : ∀ {Γ} → var Γ → ⟪ ⇧ (env Γ) ⇒ ⇧ val ⟫
lookup^♯[_] : ∀ {Γ} → var Γ → ⟪ ⇧ (env^♯ Γ) ⇒ val^♯ ⟫
```

Our calculation will be parameterized by an abstraction for ↑ ⟦_⟧$^u$:

```
postulate
  ↑⟦_⟧^u♯ : unary → ⟪ val^♯ ⇒ val^♯ ⟫
  α[⟦⟧^u] : ∀ {u v^♯} → α[ ⇄val⇄ ⇒⌜⇄^c⌝ ⇄val⇄ ] · (pure · ↑⟦ u ⟧^u) · v^♯
    ⊑ pure · ↑⟦ u ⟧^u♯ · v^♯
```

We are now set up to calculate eval$^{m\sharp}$[ $e$ ] from its specification α[ ⇄ env ⇄ ⇒ ⌜⇄$^c$⌝ ⇄ val ⇄ ] · eval$^m$[ $e$ ]. Because we want to "discover" eval$^{m\sharp}$[ $e$ ], rather than verify it a-posteriori, we state its existence and then calculate its implementation:

```
eval^m♯[_] : ∀ {Γ} → exp Γ → ⟪ ⇧ (env^♯ Γ) ⇒ val^♯ ⟫
```

We begin by stating the type of our calculation:

```
α[eval^m] : ∀ {Γ} (e : exp Γ) (ρ^♯ : ⟪ ⇧ (env^♯ Γ) ⟫)
          → α[ ⇄env⇄ ⇒⌜⇄^c⌝ ⇄val⇄ ] · eval^m[ e ] · ρ^♯
            ⊑ return · (eval^m♯[ e ] · ρ^♯)
```

and proceed by induction, the first case being numeric expressions. Each case will make use of our "proof mode" library, which we have developed in pure Agda to support calculation-style notation.

***Numeric literals*** We begin by stating the goal. We do this using our proof mode library with notation [[_]]:

```
α[eval^m] (Num n) ρ^♯ = [proof−mode]
  do [[ α[ ⇄env⇄ ⇒⌜⇄^c⌝ ⇄val⇄ ] · eval^m[ Num n ] · ρ^♯ ]]
```

This state is definitionally equal to the expansion of α[_]:

```
▸ [[ (pure · η[ ⇄val⇄ ] ⋄ eval^m[ Num n ] ⋄ γ^c[ ⇄env⇄ ]) · ρ^♯ ]]
```

Next we unfold the definition of _ ⋄ _, also by Agda computation:

```
▸ [[ (pure · η[ ⇄val⇄ ]) ∗ · (eval^m[ Num n ] ∗ · (γ^c[ ⇄env⇄ ] · ρ^♯)) ]]
```

The next step is to focus to the right of the application and replace eval$^m$[ Num $n$ ] ∗ · $R$ with its denotation return · ↑⟨ $n$ ⟩, which we prove by an auxiliary lemma $\beta$ − eval$^m$[Num]:

```
▸ [focus−right [·] of (pure · η[ ⇄val⇄ ]) ∗ ] begin
  do [[ eval^m[ Num n ] ∗ · (γ^c[ ⇄env⇄ ] · ρ^♯) ]]
   ▸ ≀ β−eval^m[Num] {R = γ^c[ ⇄env⇄ ] · ρ^♯ } ⟅[⊑]
   ▸ [[ return · ↑⟨ n ⟩ ]]
  end
▸ [[ (pure · η[ ⇄val⇄ ]) ∗ · (return · ↑⟨ n ⟩) ]]
```

Next we use the monad right-unit law to eliminate the application of an extension to a pure value:

```
▸ ≀ right−unit[∗][ pure · η[ ⇄val⇄ ] ] ⟅[≈]
▸ [[ pure · η[ ⇄val⇄ ] · ↑⟨ n ⟩ ]]
▸ [[ return · (η[ ⇄val⇄ ] · ↑⟨ n ⟩) ]]
```

It is at this point which we have reached a pure computation, absent of any specification effect. We declare this expression then to be the definition of eval$^{m\sharp}$[ Num $n$ ] and conclude:

```
▸ [[ return · (eval^m♯[ Num n ] · ρ^♯) ]] ■
```

***Variables*** The calculation for variables is more interesting, as it doesn't ignore the environment γ$^c$[ ⇄ *env* ⇄ ] · ρ$^\sharp$. We begin again by stating the goal:

```
α[eval^m] (Var x) ρ^♯ = [proof−mode]
  do [[ α[ ⇄env⇄ ⇒⌜⇄^c⌝ ⇄val⇄ ] · eval^m[ Var x ] · ρ^♯ ]]
```

As before, the first thing we do is unfold the definition of α[_]:

```
▸ [[ (pure · η[ ⇄val⇄ ] ⋄ eval^m[ Var x ] ⋄ γ^c[ ⇄env⇄ ]) · ρ^♯ ]]
▸ [[ (pure · η[ ⇄val⇄ ]) ∗ · (eval^m[ Var x ] ∗ · (γ^c[ ⇄env⇄ ] · ρ^♯)) ]]
```

Next we focus to the right of the left-most, and left of the rightmost _ ⋄ _ operator:

```
▸ [focus−right [·] of (pure · η[ ⇄val⇄ ]) ∗ ] begin
  do [[ eval^m[ Var x ] ∗ · (γ^c[ ⇄env⇄ ] · ρ^♯) ]]
   ▸ [focus−left [·] of γ^c[ ⇄env⇄ ] · ρ^♯ ] begin
     do [[ eval^m[ Var x ] ∗ ]]
```

Here we recognize that the *specification* for the semantics of Var $x$ is equivalent to the *computation* of looking up a variable in the environment, using an auxiliary proof $\beta$ − Faexp[Var]:

```
▸ [focus−in [∗] ] begin
  do [[ eval^m[ Var x ] ]]
   ▸ ≀ β−eval^m[Var] {x = x} ⟅[≈]
   ▸ [[ pure · lookup[ x ] ]]
  end
▸ [[ (pure · lookup[ x ]) ∗ ]]
end
```

Next we exploit the relationship between concrete environment lookup and abstract environment lookup: α[ ⇄ val ⇄ ⇒ ⌜⇄$^c$⌝

$\rightleftarrows \mathsf{val} \rightleftarrows\,] \cdot (\mathsf{pure} \cdot \mathsf{lookup}[\,x\,]) \sqsubseteq \mathsf{pure} \cdot \mathsf{lookup}^\sharp[\,x\,]$. To arrive at $\alpha[\,\rightleftarrows \mathsf{val} \rightleftarrows\, \Rightarrow \ulcorner\rightleftarrows^c\urcorner \rightleftarrows \mathsf{val} \rightleftarrows\,] \cdot (\mathsf{pure} \cdot \mathsf{lookup}[\,x\,])$, we first reason by extensiveness of $\rightleftarrows \mathsf{val} \rightleftarrows$:

- $[\![\ (\mathsf{pure} \cdot \mathsf{lookup}[\,x\,]) * \cdot (\gamma^c[\,\rightleftarrows\mathsf{env}\rightleftarrows\,] \cdot \rho^\sharp)\ ]\!]$
- $\wr\ \mathsf{extensive}^c[*][\,\rightleftarrows\mathsf{val}\rightleftarrows\,]$ ⟆$[\sqsubseteq]$
- $[\![\ \gamma^c[\,\rightleftarrows\mathsf{val}\rightleftarrows\,] * \cdot ((\mathsf{pure} \cdot \eta[\,\rightleftarrows\mathsf{val}\rightleftarrows\,]) * \cdot$
  $((\mathsf{pure} \cdot \mathsf{lookup}[\,x\,]) * \cdot (\gamma^c[\,\rightleftarrows\mathsf{env}\rightleftarrows\,] \cdot \rho^\sharp)))\ ]\!]$

We identify the argument to the application as $\alpha[\,\rightleftarrows \mathsf{env} \rightleftarrows\, \Rightarrow \ulcorner\rightleftarrows^c\urcorner \rightleftarrows \mathsf{val} \rightleftarrows\,] \cdot (\mathsf{pure} \cdot \mathsf{lookup}[\,x\,])$ and weaken by its abstraction:

- $[\mathsf{focus{-}right}\ [\cdot]\ \mathsf{of}\ \gamma^c[\,\rightleftarrows\mathsf{val}\rightleftarrows\,] *\,]\ \mathsf{begin}$
  $\mathsf{do}\ [\![\ (\mathsf{pure} \cdot \eta[\,\rightleftarrows\mathsf{val}\rightleftarrows\,]) * \cdot ((\mathsf{pure} \cdot \mathsf{lookup}[\,x\,]) * \cdot (\gamma^c[\,\rightleftarrows\mathsf{env}\rightleftarrows\,] \cdot \rho^\sharp))\ ]\!]$
  - $[\![\ \alpha[\,\rightleftarrows\mathsf{env}\rightleftarrows\, \Rightarrow \ulcorner\rightleftarrows^c\urcorner \rightleftarrows\mathsf{val}\rightleftarrows\,] \cdot (\mathsf{pure} \cdot \mathsf{lookup}[\,x\,]) \cdot \rho^\sharp\ ]\!]$
  - $\wr\ \alpha[\mathsf{lookup}]\ \{\mathsf{x} = x\}\ \{\rho^\sharp\}$ ⟆$[\sqsubseteq]$
  - $[\![\ \mathsf{pure} \cdot \mathsf{lookup}^\sharp[\,x\,] \cdot \rho^\sharp\ ]\!]$
  - $[\![\ \mathsf{return} \cdot (\mathsf{lookup}^\sharp[\,x\,] \cdot \rho^\sharp)\ ]\!]$

  $\mathsf{end}$
- $[\![\ \gamma^c[\,\rightleftarrows\mathsf{val}\rightleftarrows\,] * \cdot (\mathsf{return} \cdot (\mathsf{lookup}^\sharp[\,x\,] \cdot \rho^\sharp))\ ]\!]$

$\mathsf{end}$

- $[\![\ (\mathsf{pure} \cdot \eta[\,\rightleftarrows\mathsf{val}\rightleftarrows\,]) * \cdot (\gamma^c[\,\rightleftarrows\mathsf{val}\rightleftarrows\,] * \cdot (\mathsf{return} \cdot (\mathsf{lookup}^\sharp[\,x\,] \cdot \rho^\sharp)))\ ]\!]$

Finally we apply the reductive property of $\rightleftarrows \mathsf{val} \rightleftarrows$:

- $\wr\ \mathsf{reductive}^c[*][\,\rightleftarrows\mathsf{val}\rightleftarrows\,]$ ⟆$[\sqsubseteq]$
- $[\![\ \mathsf{return} \cdot (\mathsf{lookup}^\sharp[\,x\,] \cdot \rho^\sharp)\ ]\!]$

and declare the result as defining $\mathsf{eval}^\sharp[\,\mathsf{Var}\ x\,]$ and conclude:

- $[\![\ \mathsf{return} \cdot (\mathsf{eval}^{m\sharp}[\,\mathsf{Var}\ x\,] \cdot \rho^\sharp)\ ]\!]$ ■

***Unary operators*** The calculation of unary operators is interesting because it leverages an inductive hypothesis for the calculation.

$\alpha[\mathsf{eval}^m]\ (\mathsf{Unary}[\,o\,]\ e)\ \rho^\sharp\ \mathsf{with}\ \alpha[\mathsf{eval}^m]\ e\ \rho^\sharp$
$\ldots \mid IH = [\mathsf{proof{-}mode}]$
$\quad \mathsf{do}\ [\![\ \alpha[\,\rightleftarrows\mathsf{env}\rightleftarrows\, \Rightarrow \ulcorner\rightleftarrows^c\urcorner \rightleftarrows\mathsf{val}\rightleftarrows\,] \cdot \mathsf{eval}^m[\,\mathsf{Unary}[\,o\,]\ e\,] \cdot \rho^\sharp\ ]\!]$

In Agda, the with notation is a variation of let-binding which also performs dependent pattern matching refinements (although this example doesn't use dependent pattern matching). We proceed as before by expanding the definition of $\alpha[_]$.

- $[\![\ (\mathsf{pure} \cdot \eta[\,\rightleftarrows\mathsf{val}\rightleftarrows\,] \diamond \mathsf{eval}^m[\,\mathsf{Unary}[\,o\,]\ e\,] \diamond \gamma^c[\,\rightleftarrows\mathsf{env}\rightleftarrows\,]) \cdot \rho^\sharp\ ]\!]$
- $[\![\ (\mathsf{pure} \cdot \eta[\,\rightleftarrows\mathsf{val}\rightleftarrows\,]) * \cdot$
  $(\mathsf{eval}^m[\,\mathsf{Unary}[\,o\,]\ e\,] * \cdot (\gamma^c[\,\rightleftarrows\mathsf{env}\rightleftarrows\,] \cdot \rho^\sharp))\ ]\!]$

As before we focus between then $_ \diamond _$s.

- $[\mathsf{focus{-}right}\ [\cdot]\ \mathsf{of}\ (\mathsf{pure} \cdot \eta[\,\rightleftarrows\mathsf{val}\rightleftarrows\,]) *\,]\ \mathsf{begin}$
  $\mathsf{do}\ [\![\ \mathsf{eval}^m[\,\mathsf{Unary}[\,o\,]\ e\,] * \cdot (\gamma^c[\,\rightleftarrows\mathsf{env}\rightleftarrows\,] \cdot \rho^\sharp)\ ]\!]$
  - $[\mathsf{focus{-}left}\ [\cdot]\ \mathsf{of}\ \gamma^c[\,\rightleftarrows\mathsf{env}\rightleftarrows\,] \cdot \rho^\sharp\,]\ \mathsf{begin}$
    $\mathsf{do}\ [\![\ \mathsf{eval}^m[\,\mathsf{Unary}[\,o\,]\ e\,] *\ ]\!]$

We then replace the $\mathsf{eval}^m[\,\mathsf{Unary}[\,o\,]\ e\,]$ *specification* with an equivalent *computation*: $\mathsf{pure} \cdot \uparrow [\![\,o\,]\!]^u$.

- $[\mathsf{focus{-}in}\ [*]\,]\ \mathsf{begin}$
  $\mathsf{do}\ [\![\ \mathsf{eval}^m[\,\mathsf{Unary}[\,o\,]\ e\,]\ ]\!]$
  - $\wr\ \beta{-}\mathsf{eval}^m[\mathsf{Unary}]$ ⟆$[\approx]$
  - $[\![\ \mathsf{pure} \cdot \uparrow[\![\,o\,]\!]^u \diamond \mathsf{eval}^m[\,e\,]\ ]\!]$

  $\mathsf{end}$
- $[\![\ (\mathsf{pure} \cdot \uparrow[\![\,o\,]\!]^u \diamond \mathsf{eval}^m[\,e\,]) *\ ]\!]$

$\mathsf{end}$

We then reassociate.

- $[\![\ (\mathsf{pure} \cdot \uparrow[\![\,o\,]\!]^u \diamond \mathsf{eval}^m[\,e\,]) * \cdot (\gamma^c[\,\rightleftarrows\mathsf{env}\rightleftarrows\,] \cdot \rho^\sharp)\ ]\!]$
- $\wr\ \mathsf{associative}[*][\ \mathsf{pure} \cdot \uparrow[\![\,o\,]\!]^u\ ,\ \mathsf{eval}^m[\,e\,]\ ,\ \gamma^c[\,\rightleftarrows\mathsf{env}\rightleftarrows\,] \cdot \rho^\sharp\,]$ ⟆$[\approx]$
- $[\![\ (\mathsf{pure} \cdot \uparrow[\![\,o\,]\!]^u) * \cdot (\mathsf{eval}^m[\,e\,] * \cdot (\gamma^c[\,\rightleftarrows\mathsf{env}\rightleftarrows\,] \cdot \rho^\sharp))\ ]\!]$

We focus to the argument of the application and apply extensiveness of $\rightleftarrows \mathsf{val} \rightleftarrows$:

- $[\mathsf{focus{-}right}\ [\cdot]\ \mathsf{of}\ (\mathsf{pure} \cdot \uparrow[\![\,o\,]\!]^u) *\,]\ \mathsf{begin}$
  $\mathsf{do}\ [\![\ \mathsf{eval}^m[\,e\,] * \cdot (\gamma^c[\,\rightleftarrows\mathsf{env}\rightleftarrows\,] \cdot \rho^\sharp)\ ]\!]$
  - $\wr\ \mathsf{extensive}^c[*][\,\rightleftarrows\mathsf{val}\rightleftarrows\,]$ ⟆$[\sqsubseteq]$
  - $[\![\ \gamma^c[\,\rightleftarrows\mathsf{val}\rightleftarrows\,] * \cdot ((\mathsf{pure} \cdot \eta[\,\rightleftarrows\mathsf{val}\rightleftarrows\,]) * \cdot$
    $(\mathsf{eval}^m[\,e\,] * \cdot (\gamma^c[\,\rightleftarrows\mathsf{env}\rightleftarrows\,] \cdot \rho^\sharp)))\ ]\!]$

We recognize the argument to be $\alpha[\,\rightleftarrows \mathsf{env} \rightleftarrows\, \Rightarrow \ulcorner\rightleftarrows^c\urcorner \rightleftarrows \mathsf{val} \rightleftarrows\,] \cdot \mathsf{eval}^m[\,e\,] \cdot \rho^\sharp$, which we can weaken to $\mathsf{pure} \cdot \mathsf{eval}^{m\sharp}[\,e\,] \cdot \rho^\sharp$ from the inductive hypothesis:

- $[\mathsf{focus{-}right}\ [\cdot]\ \mathsf{of}\ \gamma^c[\,\rightleftarrows\mathsf{val}\rightleftarrows\,] *\,]\ \mathsf{begin}$
  $\mathsf{do}\ [\![\ (\mathsf{pure} \cdot \eta[\,\rightleftarrows\mathsf{val}\rightleftarrows\,]) * \cdot (\mathsf{eval}^m[\,e\,] * \cdot (\gamma^c[\,\rightleftarrows\mathsf{env}\rightleftarrows\,] \cdot \rho^\sharp))\ ]\!]$
  - $[\![\ \alpha[\,\rightleftarrows\mathsf{env}\rightleftarrows\, \Rightarrow \ulcorner\rightleftarrows^c\urcorner \rightleftarrows\mathsf{val}\rightleftarrows\,] \cdot \mathsf{eval}^m[\,e\,] \cdot \rho^\sharp\ ]\!]$
  - $\wr\ IH$ ⟆$[\sqsubseteq]$
  - $[\![\ \mathsf{pure} \cdot \mathsf{eval}^{m\sharp}[\,e\,] \cdot \rho^\sharp\ ]\!]$
  - $[\![\ \mathsf{return} \cdot (\mathsf{eval}^{m\sharp}[\,e\,] \cdot \rho^\sharp)\ ]\!]$

  $\mathsf{end}$
- $[\![\ \gamma^c[\,\rightleftarrows\mathsf{val}\rightleftarrows\,] * \cdot (\mathsf{return} \cdot (\mathsf{eval}^{m\sharp}[\,e\,] \cdot \rho^\sharp))\ ]\!]$

We apply the monad right unit to eliminate the extension:

  - $\wr\ \mathsf{right{-}unit}[*][\ \gamma^c[\,\rightleftarrows\mathsf{val}\rightleftarrows\,]\ ]$ ⟆$[\approx]$
  - $[\![\ \gamma^c[\,\rightleftarrows\mathsf{val}\rightleftarrows\,] \cdot (\mathsf{eval}^{m\sharp}[\,e\,] \cdot \rho^\sharp)\ ]\!]$

  $\mathsf{end}$
- $[\![\ (\mathsf{pure} \cdot \uparrow[\![\,o\,]\!]^u) * \cdot (\gamma^c[\,\rightleftarrows\mathsf{val}\rightleftarrows\,] \cdot (\mathsf{eval}^{m\sharp}[\,e\,] \cdot \rho^\sharp))\ ]\!]$

$\mathsf{end}$

Next we recognize this as an abstraction of $[\![\,u\,]\!]^u$, for which we have parameterized the calculation:

- $[\![\ (\mathsf{pure} \cdot \eta[\,\rightleftarrows\mathsf{val}\rightleftarrows\,]) * \cdot$
  $((\mathsf{pure} \cdot \uparrow[\![\,o\,]\!]^u) * \cdot (\gamma^c[\,\rightleftarrows\mathsf{val}\rightleftarrows\,] \cdot (\mathsf{eval}^{m\sharp}[\,e\,] \cdot \rho^\sharp)))\ ]\!]$
- $[\![\ \alpha[\,\rightleftarrows\mathsf{val}\rightleftarrows\, \Rightarrow \ulcorner\rightleftarrows^c\urcorner \rightleftarrows\mathsf{val}\rightleftarrows\,] \cdot (\mathsf{pure} \cdot \uparrow[\![\,o\,]\!]^u) \cdot (\mathsf{eval}^{m\sharp}[\,e\,] \cdot \rho^\sharp)\ ]\!]$
- $\wr\ \alpha[[\![\,]\!]^u]$ ⟆$[\sqsubseteq]$
- $[\![\ \mathsf{pure} \cdot \uparrow[\![\,o\,]\!]^{u\sharp} \cdot (\mathsf{eval}^{m\sharp}[\,e\,] \cdot \rho^\sharp)\ ]\!]$
- $[\![\ \mathsf{return} \cdot (\uparrow[\![\,o\,]\!]^{u\sharp} \cdot (\mathsf{eval}^{m\sharp}[\,e\,] \cdot \rho^\sharp))\ ]\!]$

We declare the result to be the definition of $\mathsf{eval}^\sharp$ and conclude:

- $[\![\ \mathsf{return} \cdot (\mathsf{eval}^{m\sharp}[\,\mathsf{Unary}[\,o\,]\ e\,] \cdot \rho^\sharp)\ ]\!]$ ■

We can then define $\mathsf{eval}^{m\sharp}$ as the result of calculation:

$\mathsf{eval}^{m\sharp}[\,\mathsf{Num}\ n\,] = \mathsf{mk}[\Rightarrow]\ (\lambda\ \rho^\sharp \rightarrow \eta[\,\rightleftarrows\mathbb{Z}\rightleftarrows\,] \cdot \uparrow\langle\, n\,\rangle)$
$\mathsf{eval}^{m\sharp}[\,\mathsf{Var}\ x\,] = \mathsf{mk}[\Rightarrow]\ (\lambda\ \rho^\sharp \rightarrow \mathsf{lookup}^\sharp[\,x\,] \cdot \rho^\sharp)$
$\mathsf{eval}^{m\sharp}[\,\mathsf{Unary}[\,u\,]\ e\,] = \mathsf{mk}[\Rightarrow]\ (\lambda\ \rho^\sharp \rightarrow \uparrow[\![\,u\,]\!]^{u\sharp} \cdot (\mathsf{eval}^{m\sharp}[\,e\,] \cdot \rho^\sharp))$

### 6.4 End to End: Connection to the Collecting Semantics

Recall that the original collecting semantics we wish to abstract, *eval*, is the extension of the monadic semantics, $eval^m*$. To establish the final proof of abstraction, we promote the partial order of the previous section between monadic functions:

```
α[eval] : α[ ⇄env⇄ ⇒⌜⇄^c⌝ ⇄val⇄ ] · eval^m[ e ]
          ⊑ pure · eval^m♯[ e ]
```

to a partial ordering between extended functions:

```
α[eval*] : α[ (⇄env⇄ *⌜⇄^c⌝) ⇒⌜⇄⌝ (⇄val⇄ *⌜⇄^c⌝) ] · eval^m[ e ] *
           ⊑ (pure · eval^m♯[ e ]) *
```

where `_*⌜⇄^c⌝` is the promotion operator from Kleisli to classical Galois connections, and `_⇒⌜⇄⌝_` is the standard classical Galois connection pointwise lifting operator.

We define `_*⌜⇄^c⌝` following the proof of inclusion from Kleisli Galois connections into classical Galois connections:

```
_*⌜⇄^c⌝ : ∀ {A₁ A₂ : POSet} → A₁ ⇄^c A₂ → (℘ A₁ ⇄ ℘ A₂)
⇄A⇄ *⌜⇄^c⌝ = record
  { α[_] = (pure · η[ ⇄A⇄ ]) *
  ; γ[_] = γ^c[ ⇄A⇄ ] *
  ; extensive[_] = … ; reductive[_] = … }
```

and we prove soundness and completeness following the definitions given in section 7:

```
sound/complete :
  ∀ {A₁ A₂ B₁ B₂ : POSet}
    (⇄A⇄ : A₁ ⇄^c A₂) (⇄B⇄ : B₁ ⇄^c B₂)
    (f : ⟪ A₁ ⇒ ℘ B₁ ⟫) (f♯ : ⟪ A₂ ⇒ ℘ B₂ ⟫)
  → (∀ x♯ →  α[ ⇄A⇄ ⇒⌜⇄^c⌝ ⇄B⇄ ] · f · x♯
              ⊑ f♯ · x♯)
  ↔ (∀ X♯ →  α[ ⇄A⇄ *⌜⇄^c⌝ ⇒⌜⇄⌝ ⇄B⇄ *⌜⇄^c⌝ ] · f * · X♯
              ⊑ f♯ * · X♯)
sound/complete = …
```

*α*[*eval*∗] is then defined as an application of the soundness direction of *sound*/*complete*:

```
α[eval^m*] : ∀ {Γ} (e : exp Γ) (R♯ : ⟪ ℘ (⇧ (env♯ Γ)) ⟫)
  → α[ (⇄env⇄ *⌜⇄^c⌝) ⇒⌜⇄⌝ (⇄val⇄ *⌜⇄^c⌝) ] · eval^m[ e ] * · R♯
  ⊑ (pure · eval^m♯[ e ]) * · R♯
α[eval^m*] e R♯ =
  π₁ (sound/complete ⇄env⇄ ⇄val⇄ eval^m[ e ] (pure · eval^m♯[ e ]))
  (α[eval^m] e) R♯
```

The next section describes the soundness and completeness result which we rely on in this section in more detail, and develops the foundations of Kleisli Galois connections.

## 7. Foundations of Kleisli Galois Connections

Kleisli Galois connections are formed by re-targeting the classical Galois connection framework from the category of posets to the powerset Kleisli category, where morphisms are monotonic monadic functions, as described in section 4.2.

Unfolding the definition of $extensive^m$ and $reductive^m$ from section 4.2 we reveal equivalent, more intuitive properties, which we call $soundness^m$ and $completeness^m$:

$$soundness^m : \forall(x).\exists(y).y \in \alpha^m(x) \land x \in \gamma^m(y)$$

$$completeness^m : \forall(x_1^\sharp, x_2^\sharp, x).x \in \gamma^m(x_1^\sharp) \land x_2^\sharp \in \alpha^m(x) \to x_2^\sharp \sqsubseteq x_1^\sharp$$

These definitions provide a *relational* setup for the soundness and completeness of $\alpha^m$ and $\gamma^m$. In fact, the model for the monadic space $A \to \wp(B)$ is precisely $A \nrightarrow B \nrightarrow Set$,[1] i.e. monotonic relations over $A$ and $B$. We have therefore recovered a relational setting for soundness and completeness of abstractions between sets, purely by following the natural consequences of instantiating the Galois connection framework to the powerset Kleisli category.

[1] Here $\nrightarrow$ denotes antitonic functions; *Set* is ordered by implication.

### 7.1 Lifting Kleisli Galois Connections

The final step of our calculational relies on bridging the gap between Kleisli and classical Galois connections. This bridge enables us to construct a Kleisli Galois connection

$$env \to \wp(val) \underset{\alpha^{e \to^m v}}{\overset{\gamma^{e \to^m v}}{\leftrightarrows}} env^\sharp \to \wp(val^\sharp)$$

and calculation $\alpha^{e \to^m v}(eval^m[e]) \sqsubseteq eval^{m\sharp}[e]$. and lift both systematically to classical results, and to do so without any loss of generality. We formalize these notions in the following theorems:

**Theorem 1** (GC-Soundness).✓ *For every Kleisli Galois connection*

$$A \underset{\alpha^m}{\overset{\gamma^m}{\leftrightarrows}} B$$

*there exists a classical Galois connection*

$$\wp(A) \underset{\alpha*}{\overset{\gamma*}{\leftrightarrows}} \wp(B)$$

*where* $\alpha* := \alpha^m*$ *and* $\gamma* := \gamma^m*$.

**Theorem 2** (GC-Completeness).✓ *For every classical Galois connection*

$$\wp(A) \underset{\alpha}{\overset{\gamma}{\leftrightarrows}} \wp(B)$$

*where* $\alpha$ *and* $\gamma$ *are of the form* $\alpha = \alpha^m*$ *and* $\gamma = \gamma^m*$, *there exists a Kleisli Galois connection*

$$A \underset{\alpha^m}{\overset{\gamma^m}{\leftrightarrows}} B$$

Next we show how to lift Kleisli Galois connections pointwise to a classical Galois connection over extensions:

**Lemma 8** (Pointwise-lifting-extensions).✓ *Given Kleisli Galois connections*

$$A_1 \underset{\alpha^{mA}}{\overset{\gamma^{mA}}{\leftrightarrows}} A_2 \quad B_1 \underset{\alpha^{mB}}{\overset{\gamma^{mB}}{\leftrightarrows}} B_2$$

*there exists a classical Galois connection*

$$\wp(A_1) \nrightarrow \wp(B_1) \underset{\alpha^{(A \to^m B)*}}{\overset{\gamma^{(A \to^m B)*}}{\leftrightarrows}} \wp(A_2) \nrightarrow \wp(B_2)$$

*where*

$$\alpha^{(A \to^m B)*}(F) := \alpha^{mB} * \circ F \circ \gamma^{mA}* \quad \gamma^{(A \to^m B)*}(F^\sharp) := \gamma^{mB} * \circ F^\sharp \circ \gamma^{mA}*$$

And finally we establish an isomorphism of partial ordering between monadic functions and their extensions:

**Theorem 3** (Soundness).✓ *Given Kleisli Galois connections*

$$A_1 \underset{\alpha^{mA}}{\overset{\gamma^{mA}}{\leftrightarrows}} A_2 \quad B_1 \underset{\alpha^{mB}}{\overset{\gamma^{mB}}{\leftrightarrows}} B_2$$

*and functions* $f \in A_1 \nrightarrow \wp(B_1)$ *and* $f^\sharp \in A_2 \nrightarrow \wp(B_2)$, *partial orders under the Kleisli pointwise lifting imply partial orders under extension:*

$$\alpha^{A \to^m B}(f) \sqsubseteq f^\sharp \Rightarrow \alpha^{(A \to^m B)*}(f*) \sqsubseteq f^\sharp * .$$

**Theorem 4** (Completeness).✓ *Given Kleisli Galois connections*

$$A_1 \underset{\alpha^{mA}}{\overset{\gamma^{mA}}{\leftrightarrows}} A_2 \quad B_1 \underset{\alpha^{mB}}{\overset{\gamma^{mB}}{\leftrightarrows}} B_2$$

*and functions* $f \in A_1 \nrightarrow \wp(B_1)$ *and* $f^\sharp \in A_2 \nrightarrow \wp(B_2)$, *partial orders under the Kleisli pointwise lifting for extensions imply partial orders without extension:*

$$\alpha^{(A \to^m B)*}(f*) \sqsubseteq f^\sharp* \Rightarrow \alpha^{A \to^m B}(f) \sqsubseteq f^\sharp.$$

### 7.2 Constructive Galois Connections

Analogously to Kleisli Galois connection, we state extensiveness and reductiveness as equivalent soundness and completeness properties:

$$soundness^c : \forall(x).x \in \gamma(\eta(x))$$
$$completeness^c : \forall(x^\sharp, x).x \in \gamma(x^\sharp) \Rightarrow \eta(x) \sqsubseteq x^\sharp$$

These statements have even stronger intuitive meaning that that of Kleisli Galois connections. *soundness*$^c$ states that $x$ must be in the concretization of its abstraction, and *completeness*$^c$ states that the best abstraction for $x$, i.e. $\eta(x)$, must be better any other abstraction for $x$, i.e. $x^\sharp$.

Constructive Galois connections are initially motivated by the need for pure abstraction functions during the process of calculation, and simultaneously from the observation that abstraction functions are often pure function in practice. What is surprising is that constructive Galois connections are not a special case of Kleisli Galois connections: *all* Kleisli Galois connections are constructive.

**Theorem 5.**✓ *The set of Kleisli Galois connections is isomorphic to the set of constructive Galois connections.*

*Proof.* The easy direction is constructing a Kleisli Galois connection from a constructive Galois connection. Given a constructive Galois connection $A \underset{\eta}{\overset{\gamma^c}{\leftrightarrows}} B$, we construct the following Kleisli Galois connection:

$$\alpha^m : A \to \wp(B) \qquad \gamma^m : B \to \wp(A)$$
$$\alpha^m = pure(\eta) \qquad \gamma^m = \gamma^c$$

Proofs for extensiveness and reductiveness follow definitionally.

The next step is to construct a Constructive Galois connection from a Kleisli Galois connection $A \underset{\alpha^m}{\overset{\gamma^m}{\leftrightarrows}} B$. This at first seems paradoxical, since it requires constructing an abstraction *function* $\eta : A \to B$ from the given abstraction *specification* $\alpha^m : A \to \wp(B)$. However, we are able exploit the property of *soundness*$^m$, which is equivalent to *extensive*$^m$, from the definition of Kleisli Galois connections to define $\eta$.

Recall the soundness judgement for Kleisli Galois connections, which arise directly from the definition of *return* and _*.

$$soundness^m : \forall(x).\exists(y).y \in \alpha(x) \wedge x \in \gamma(y)$$

Given a proof of *soundness*$^m$, we use the axiom of choice to extract the existentially quantified $y$ given an $x$. In fact, the axiom of choice is not an axiom in constructive logic, rather it is a *theorem* of choice, which can be written in Agda.

```
choice : ∀ {A B} {P : A → B → Set} → (∀ x → ∃ y st P x y) → A → B
choice f x with f x
... | ∃ y ,, P[x,y] = y
```

Using the axiom of choice we easily define $\eta$ and $\gamma^c$.

$$\eta \in A \to B \qquad \gamma^c \in B \to \wp(A)$$
$$\eta(x) = y \text{ where } (\exists y : y \in \alpha^m(x) \wedge x \in \gamma^m(y)) \qquad \gamma^c = \gamma^m$$

In order for $\eta$ and $\gamma^c$ to be a valid Galois connection we must still prove extensiveness and reductiveness. To do so we instead prove *soundness*$^c$ and *completeness*$^c$, which are equivalent to *extensive*$^c$ and *reductive*$^c$. These proofs follow from the soundness evidence attached to $\eta(x)$ and its use of the axiom of choice.

**Lemma 9** (*soundness*$^c$)**.**✓ $\forall(x).x \in \gamma^c(\eta(x))$.

**Lemma 10** (*completeness*$^c$)**.**✓ $\forall(x^\sharp, x).x \in \gamma^c(x^\sharp) \to \eta(x) \sqsubseteq x^\sharp$.

Finally, to establish the isomorphism, we show that transforming a Kleisli Galois connection into a constructive one and back results in the same Galois connection. To show this we apply the following lemma, a restatement of its classical analogue [19, p.239] in the Kleisli setting:

**Lemma 11** (Kleisli-Uniqueness)**.**✓ *Given two Kleisli Galois connections* $A \underset{\alpha_1^m}{\overset{\gamma_1^m}{\leftrightarrows}} B$ *and* $A \underset{\alpha_2^m}{\overset{\gamma_2^m}{\leftrightarrows}} B$, $\alpha_1^m = \alpha_2^m$ *if and only if* $\gamma_1^m = \gamma_2^m$

To use this lemma, we recognize that the concretization functions $\gamma^m$ and $\gamma^c$ are definitionally the same for both mappings between Kleisli and constructive Galois connections. It then follows that $\alpha^m$ and $pure(\eta)$ must be equal.

□

The consequence of the isomorphism between Kleisli and constructive Galois connections is that we may work directly with constructive Galois connections without any loss of generality. Furthermore, we can assume a pure "extraction function" $\eta$ for every Kleisli abstraction function $\alpha^m$ where $\alpha^m = pure(\eta)$.

Finally, our proof of isomorphism gives a foundational explanation for *why* some Galois connections happen to have fully computational functions as their abstraction functions. These pure abstraction functions are no accident; they are induced by the Kleisli Galois connection setup embedded in constructive logic, where the axiom of choice is definable as a theorem with computational content.

## 8. Related Work

This work connects two long strands of research: abstract interpretation and dependently typed programming. The former is founded on the pioneering work of Cousot and Cousot [9, 10]; the latter on that of Martin-Löf [15], embodied in Norell's Agda [20]. A key technical insight of ours is to use a monadic structure for Galois connections and proofs by calculus, following the example of Moggi [18] for the $\lambda$-calculus.

***Calculational abstract interpretation*** Cousot describes the calculation approach to abstract interpretation by example in his lecture notes [8], the foundations for which can be found in [7], and recently introduced a unifying calculus for Galois connections [12]. Other notable uses of calculational abstract interpretation include the calculational derivation of higher order control flow analysis [16] and the calculation of polynomial time graph algorithms [23].

Our work mechanizes Cousot's calculations, and provides a suitable foundation for mechanizing other instances of calculational abstract interpretation.

***Calculational program design*** Related to the calculation of abstract interpreters is the calculation of programs, long advocated by Bird and others as calculational program design [2, 3].

Calculational program design has been successfully mechanized in proof assistants [26]. This practice does not encounter the non-constructive metatheory issues which show up in mechanizing calculational abstract interpreters. In mechanized calculational program design, specifications are fully constructive, whose inhabitants can be readily executed as programs. In abstract interpretations the specifications are inherently non-constructive, which leads to the need for new theoretical machinery.

***Verified static analyses*** Verified abstract interpretation has seen many promising results [1, 4, 6, 22], scaling up recently to large-scale real-world static analyzers [14]. Mechanized abstract interpretation has yet to benefit from being built on a solid, compositional Galois connection framework. Until now approach have used either "α-only" or "γ-only" encodings of soundness and (sometimes) completeness. Our techniques for isolating specification effects should readily apply to these existing approaches.

***Monadic abstract interpretation*** The use of monads in abstract interpretation has recently been used to good effect [13, 24]. However that work uses monads to structure the language semantics, whereas our approach has been to use monadic structure in the Galois connections and proofs by calculus.

***Galculator*** The *Galculator* [25] is a proof assistant founded on an algebra of Galois connections. This tool is similar to ours in that it mechanically verifies Galois connection calculations; additionally it fully automates the calculational derivations themselves. Our approach is more general, supporting arbitrary set-theoretic reasoning and embeded within a general purpose proof assistant, however their approach is fully automated for the small set of derivations which reside within their supported theory. We foresee a marriage of the two approaches, where simple algebraic calculations are derived automatically, yet more complicated connections are still expressible and provable within the same mechanized framework.

***Future directions*** Now that we have established a foundation for constructive Galois connection calculation, we see value in verifying larger derivations (e.g. [17, 23]). Furthermore we would like to explore whether or not our techniques have any benefit in the space of general-purpose program calculations *à la* Bird.

There have also been recent developments on compositional abstract interpretation frameworks [13] where abstract interpreter implementations and their proofs of soundness via Galois connection are systematically derived side-by-side. Their framework relies on correctness properties transported by *Galois transformers*, which we believe would greatly benefit from mechanization, because they hold both computational and specification content.

## 9. Conclusions

Over fifteen years ago, when concluding "The calculational design of a generic abstract interpreter" [7, p. 85], Cousot wrote:

> The emphasis in these notes has been on the correctness of the design by calculus. The mechanized verification of this formal development using a proof assistant can be foreseen with automatic extraction of a correct program from its correctness proof.

This paper realizes that vision, giving the first mechanically verified proof of correctness for Cousot's abstract interpreter. Our proof "by calculus" closely follows the original paper-and-pencil proof. The primary discrepancy being the use of monadic reasoning to isolate *specification effects*. By maintaining this monadic discipline, we are able to verify calculations by Galois connections *and* extract computation content from pure results. The resulting static analyzer is correct by verified construction and therefore does not suffer from bugs present in the original.[2]

## Acknowledgments

We gratefully acknowledge the Colony Club in Washington, DC for providing a fruitful environment in which to do this research.

[2] `http://www.di.ens.fr/~cousot/aisoftware/Marktoberdorf98/Bug_History`

## References

[1] G. Barthe, D. Pichardie, and T. Rezk. A certified lightweight non-interference java bytecode verifier. In R. D. Nicola, editor, *Programming Languages and Systems*, Lecture Notes in Computer Science. Springer Berlin Heidelberg, 2007.

[2] R. Bird and O. de Moor. *The Algebra of Programming*. Prentice Hall, 1996.

[3] R. S. Bird. A calculus of functions for program derivation. In D. A. Turner, editor, *Research Topics in Functional Programming*. Addison-Wesley, 1990. Also available as Technical Monograph PRG-64, from the Programming Research Group, Oxford University.

[4] S. Blazy, V. Laporte, A. Maroneze, and D. Pichardie. Formal verification of a C value analysis based on abstract interpretation. In F. Logozzo and M. Fähndrich, editors, *Static Analysis*, Lecture Notes in Computer Science. Springer Berlin Heidelberg, 2013.

[5] A. Bove, P. Dybjer, and U. Norell. A brief overview of agda – A functional language with dependent types. In S. Berghofer, T. Nipkow, C. Urban, and M. Wenzel, editors, *Theorem Proving in Higher Order Logics*, Lecture Notes in Computer Science, chapter 6. Springer Berlin Heidelberg, 2009.

[6] D. Cachera and D. Pichardie. A certified denotational abstract interpreter. In M. Kaufmann and L. Paulson, editors, *Interactive Theorem Proving*, Lecture Notes in Computer Science. Springer Berlin Heidelberg, 2010.

[7] P. Cousot. The calculational design of a generic abstract interpreter. In M. Broy and R. Steinbrüggen, editors, *Calculational System Design*. NATO ASI Series F. IOS Press, Amsterdam, 1999.

[8] P. Cousot. MIT 16.399: Abstract interpretation, 2005.

[9] P. Cousot and R. Cousot. Abstract interpretation: a unified lattice model for static analysis of programs by construction or approximation of fixpoints. In *Proceedings of the 4th ACM SIGACT-SIGPLAN Symposium on Principles of Programming Languages*. ACM, 1977.

[10] P. Cousot and R. Cousot. Systematic design of program analysis frameworks. In *Proceedings of the 6th ACM SIGACT-SIGPLAN Symposium on Principles of Programming Languages*, POPL '79. ACM, 1979.

[11] P. Cousot and R. Cousot. *Basic Concepts of Abstract Interpretation*, pages 359–366. Kluwer Academic Publishers, 2004.

[12] P. Cousot and R. Cousot. A Galois connection calculus for abstract interpretation. In *Proceedings of the 41st ACM SIGPLAN-SIGACT Symposium on Principles of Programming Languages*, POPL '14. ACM, 2014.

[13] D. Darais, M. Might, and D. Van Horn. Galois transformers and modular abstract interpreters. *CoRR*, abs/1411.3962, 2014. URL `http://arxiv.org/abs/1411.3962`.

[14] J. H. Jourdan, V. Laporte, S. Blazy, X. Leroy, and D. Pichardie. A Formally-Verified C static analyzer. In *Proceedings of the 42Nd Annual ACM SIGPLAN-SIGACT Symposium on Principles of Programming Languages*, POPL '15. ACM, 2015.

[15] P. Martin-Löf. *Intuitionistic Type Theory*. Bibliopolis, 1984.

[16] J. Midtgaard and T. Jensen. A calculational approach to Control-Flow analysis by abstract interpretation. In M. Alpuente and G. Vidal, editors, *Static Analysis*, Lecture Notes in Computer Science, chapter 23. Springer Berlin Heidelberg, 2008.

[17] J. Midtgaard and T. Jensen. A calculational approach to Control-Flow analysis by abstract interpretation. In M. Alpuente and G. Vidal, editors, *SAS*, LNCS. Springer, 2008.

[18] E. Moggi. An abstract view of programming languages. Technical report, Edinburgh University, 1989.

[19] F. Nielson, H. R. Nielson, and C. Hankin. *Principles of Program Analysis*. Springer-Verlag, 1999.

[20] U. Norell. *Towards a practical programming language based on dependent type theory*. PhD thesis, Department of Computer Science and Engineering, Chalmers University of Technology, SE-412 96 Göteborg, Sweden, September 2007.

[21] U. Norell. Dependently typed programming in agda. In P. Koopman, R. Plasmeijer, and D. Swierstra, editors, *Advanced Functional Pro-*

*gramming*, Lecture Notes in Computer Science, chapter 5. Springer Berlin Heidelberg, 2009.

[22] D. Pichardie. *Interprétation abstraite en logique intuitionniste: extraction d'analyseurs Java certifiés*. PhD thesis, Université Rennes, 2005.

[23] I. Sergey, J. Midtgaard, and D. Clarke. Calculating graph algorithms for dominance and shortest path. In J. Gibbons and P. Nogueira, editors, *Mathematics of Program Construction*, Lecture Notes in Computer Science. Springer Berlin Heidelberg, 2012.

[24] I. Sergey, D. Devriese, M. Might, J. Midtgaard, D. Darais, D. Clarke, and F. Piessens. Monadic abstract interpreters. In *Proceedings of the 34th ACM SIGPLAN Conference on Programming Language Design and Implementation*. ACM, 2013.

[25] P. F. Silva and J. N. Oliveira. 'Galculator': Functional prototype of a Galois-connection based proof assistant. In *Proceedings of the 10th International ACM SIGPLAN Conference on Principles and Practice of Declarative Programming*, PPDP '08. ACM, 2008.

[26] J. Tesson, H. Hashimoto, Z. Hu, F. Loulergue, and M. Takeichi. Program calculation in Coq. In M. Johnson and D. Pavlovic, editors, *Algebraic Methodology and Software Technology*, Lecture Notes in Computer Science, chapter 10. Springer Berlin Heidelberg, 2011.